\documentclass[manuscript,twocolumn]{aa}
\usepackage{multicol}
\usepackage{graphicx}
\usepackage{txfonts}
\usepackage{epsfig}
\usepackage{times}
\usepackage{natbib}
\usepackage{url}
\usepackage{color}
\usepackage{amssymb,amsmath}
\usepackage{enumerate}
\usepackage[normalem]{ulem}
\usepackage{cancel}
\usepackage{soul}
\usepackage[
  breaklinks = true,
   colorlinks = true,
  urlcolor   = blue,
  citecolor  = blue,
  linkcolor  = blue,
]{hyperref}
\graphicspath{{./figures/}}

\newcommand{\lthread}{{l_{\rm t}}}

\newcommand{\Mm}{{\mathrm{\, Mm}}}
\newcommand{\kms}{{\mathrm{\, km \; s^{-1}}}}
\newcommand{\kmss}{{\mathrm{\, km \; s^{-2}}}}

\newcommand{\mins}{{\mathrm{\, minutes}}}
\newcommand{\G}{{\mathrm{\, G}}}
\newcommand{\ha}{{H$\alpha$ }}
\newcommand{\grav}{{{g_0}}}
\newcommand{\speedjettwo}{{235 \kms}}
\newcommand{\speedjetone}{{151 \kms}}
\newcommand{\periodlalosecondphasenu}{{72 \pm 2}}
\newcommand{\periodlalosecondphase}{{\periodlalosecondphasenu \mins}}
\newcommand{\dampingtimelalosecondphasenu}{{196 \pm 71}}
\newcommand{\dampingtimelalosecondphase}{{\dampingtimelalosecondphasenu \mins}}
\newcommand{\velocitylalosecondphasenu}{{73 \pm 16}}
\newcommand{\velocitylalosecondphase}{{\velocitylalosecondphasenu \kms}}

\newcommand{\amplitudelalosecondphasenu}{{57 \pm 15}}
\newcommand{\amplitudelalosecondphase}{{\amplitudelalosecondphasenu \Mm}}

\newcommand{\radiuslalosecondphasenu}{{159 \pm 11}}
\newcommand{\radiuslalosecondphase}{{\radiuslalosecondphasenu \Mm}}
\newcommand{\magneticlalosecondphasenu}{{22 \pm 11}}
\newcommand{\magneticlalosecondphase}{{\magneticlalosecondphasenu \G}}

\newcommand{\periodlalofirstphaseDFnu}{{69 \pm 3}}
\newcommand{\dampingtimelalofirstphaseDFnu}{{297 \pm 369}}
\newcommand{\velocitylalofirstphaseDFnu}{{24 \pm 6}}
\newcommand{\radiuslalofirstphaseDFnu}{{143 \pm 15}}
\newcommand{\magneticlalofirstphaseDFnu}{{21 \pm 11}}

\newcommand{\periodlalofirstphaseDF}{{\periodlalofirstphaseDFnu \mins}}
\newcommand{\dampingtimelalofirstphaseDF}{{\dampingtimelalofirstphaseDFnu \mins}}

\newcommand{\radiuslalofirstphaseDF}{{\radiuslalofirstphaseDFnu \Mm}}
\newcommand{\magneticlalofirstphaseDF}{{\magneticlalofirstphaseDFnu \G}}

\newcommand{\periodlalofirstphaseRFnu}{{62 \pm 2}}
\newcommand{\dampingtimelalofirstphaseRFnu}{{281 \pm 166}}
\newcommand{\velocitylalofirstphaseRFnu}{{33 \pm 3}}
\newcommand{\radiuslalofirstphaseRFnu}{{111 \pm 8}}
\newcommand{\magneticlalofirstphaseRFnu}{{19 \pm 10}}

\newcommand{\periodlalofirstphaseRF}{{\periodlalofirstphaseRFnu \mins}}
\newcommand{\dampingtimelalofirstphaseRF}{{\dampingtimelalofirstphaseRFnu \mins}}
\newcommand{\velocitylalofirstphaseRF}{{\velocitylalofirstphaseRFnu \kms}}
\newcommand{\radiuslalofirstphaseRF}{{\radiuslalofirstphaseRFnu \Mm}}
\newcommand{\magneticlalofirstphaseRF}{{\magneticlalofirstphaseRFnu \G}}

\newcommand{\rsun}{R_\odot}
\newcommand{\psun}{P_\odot}
\definecolor{fmi}{rgb}{0.89, 0.0, 0.13}

\newcommand{\fminote}[1]{\emph{\textcolor{fmi}{[[#1]]}}}
\def\sizefactorhalf{0.5} %nominal value 0.5
\def\sizefactorfull{0.9} %nominal value 1
\begin{document}
%\title{What are the causes of large amplitude oscillations in solar filaments?%
\title{Study of the excitation of large amplitude oscillations in a prominence by nearby flares}

%St. Patrick’s Day Storm 
\author{Manuel Luna
\inst{1,2},
Reetika Joshi
\inst{3,4},
Brigitte Schmieder
\inst{5,6,7},
Fernando Moreno-Insertis
\inst{8,9},
Valeriia Liakh
\inst{6}
\&
Jaume Terradas
\inst{1,2}
%EEEE
%\inst{10}
}
\institute{Departament F{\'i}sica, Universitat de les Illes Balears (UIB), E-07122 Palma de Mallorca, Spain\\ 
\email{manuel.luna@uib.es}
\and
Institute of Applied Computing \& Community Code (IAC$^3$), UIB, Spain
\and
Institute of Theoretical Astrophysics, University of Oslo, P.O. Box 1029 Blindern, N-0315 Oslo, Norway
\and
Rosseland Centre for Solar Physics, University of Oslo, P.O. Box 1029 Blindern, N-0315 Oslo, Norway
\and
LESIA, Observatoire de Paris,  Universit\'e PSL, CNRS, Sorbonne Universit\'e,  Universit\'e de Paris, 5 place Janssen, 92290  Meudon Principal Cedex, France
\and
Centre for mathematical Plasma Astrophysics, Dept. of Mathematics, KU Leuven, 3001 Leuven, Belgium
 \and
SUPA, School of Physics and Astronomy, University of Glasgow, Glasgow G12 8QQ, UK
\and
Instituto de Astrof\'{\i}sica de Canarias, E-38200 La Laguna, Tenerife, Spain 
\and
Departamento de Astrof\'{\i}sica, Universidad de La Laguna, E-38206 La Laguna, Tenerife, Spain 
}

\authorrunning{Luna et al.} 
   \titlerunning{Causes of large amplitude oscillations in a solar filament  transverse or longitudinal}

\date{Accepted October, 2024}
   
\abstract
% context heading (optional)
{Large amplitude oscillations are a common occurrence in solar prominences. These oscillations are triggered by energetic phenomena such as jets and flares. On March 14-15, 2015, a filament partially erupted in two stages, leading to oscillations in different parts of the filament.}
% aims heading (mandatory)
{In this study, we aim to explore the longitudinal oscillations resulting from the eruption, with special focus on unravelling the underlying mechanisms responsible for their initiation. We pay special attention to the huge oscillation on {March 15th.}}
% methods heading (mandatory)
{ 
The oscillations and jets are analyzed by using the time-distance technique. For the study of 
flares and their interaction with the filament, we analyse in detail the different AIA channels and use the DEM technique. 
}
%  % results heading (mandatory)
{In the initial phase of the event, a jet induces the fragmentation of the filament, which causes it to split into two segments. One of the segments remains in the same position, while the other is detached and moves to a different location.
 This causes oscillations in both segments: (a) the change of position apparently causes the detached segment to oscillate longitudinally with a period of $\periodlalofirstphaseDF$; (b) the jet flows reach the remaining filament also producing longitudinal oscillations with a period of $\periodlalofirstphaseRF$.
In the second phase, on {March 15th}, another jet seemingly activates the detached filament eruption.
After the eruption, there is an associated flare. A large longitudinal oscillation is produced in the remnant segment with a period of 
$\periodlalosecondphase$ and 
velocity amplitude $\velocitylalosecondphase$. 
During the triggering of the oscillation, bright field lines connect the flare with the filament. These only appear in the AIA 131\AA\ and 94\AA\ channels, indicating that they contain very hot plasma. The DEM analysis also confirms this result. Both indicate that a plasma around 10 MK pushes the prominence from its south-eastern side, displacing it along the field lines and initiating the oscillation.
From this evidence, the flare and not the preceding jet initiates the oscillation. The hot plasma from the flare escapes and flows into the filament channel structure.
%
%so this part will be removerd?
}
% % conclusions heading (optional)
{In this paper we shed light on how flares can initiate the huge oscillations in filaments. We propose an explanation in which part of the post-flare loops reconnect with the filament channel magnetic field lines that hosts the prominence.}

\maketitle

\section{Introduction}\label{sec:introduction}
Solar prominences (also called filaments) are highly dynamic structures that are subject to a large variety of oscillations. These oscillations can be categorised %divided 
into two main groups, namely large amplitude oscillations (LAOs) and small amplitude oscillations. 
LAOs are global oscillations where a large part of the filament participates in the oscillation, with velocities typically greater than 10$\kms$ \citep{Arregue2018}. In some cases, the velocity amplitude is close to 100$\kms$ \citep{jing2006, Luna_Large-amplitude_2017}. 
These LAOs are expected to hold a huge amount of energy due to the combination of the large mass and large velocities involved. \citet{luna_observations_2014} reported a LAO and estimated that the energy involved in the oscillation was in the range $10^{24}-10^{27}$ erg. This is in the energy range of a microflare \citep{shibata_solar_2011}. However, considering that only part of the energy is transmitted to the filament by the triggering agent, the energy needed to produce these LAOs must be enormous.
\citet{zhang_parametric_2013} conducted numerical experiments inducing LAO oscillations through impulsive heating at a single footpoint of a loop. In their simulations, the energy associated with the oscillations of the thread amounts to merely 4\% of the impulsive energy release.
In this paper, we focus on LAOs, trying to shed light on the possible mechanisms that trigger these oscillations.

Early observations of prominence oscillations have already revealed a possible relationship with flares \citep{dodson1949,bruzek1957,becker1958}. It was suggested
that their exciter was a wave, caused by the flare, which disturbs the filament located far away from the flare site and induces damped oscillations. 
\cite{Moreton1960b} confirmed that wave disturbances initiated during the impulsive phase of flares were responsible for triggering prominence oscillations far from the flare. The waves identified by the authors are nowadays known as Moreton waves. \citet{Hyder1966} observed eleven different cases where flares produced oscillations in distant prominences. More recently, thanks to both space- and ground-based instruments, observations of filament LAOs have become common. The exciters identified thus far include Moreton or EIT waves \citep{eto2002,okamoto2004,gilbert2008,asai2012,Pant2016}, EUV waves \citep{hershaw2011,liu2012, shen2014a,xue2014,takahashi2015, Devi2022}, shock waves \citep{shen2014b}, nearby jets \citep{luna_observations_2014,zhang2017,joshi_interaction_2023}, subflares and flares \citep{jing2003, jing2006,vrsnak2007,Li2012}, and filament eruptions 
\citep{isobe2006,isobe2007, pouget07,chen2008,foullon2009,bocchialini2011,Luna_Large-amplitude_2017,mazumder_simultaneous_2020}. In \citet{luna_gong_2018} 196 oscillations are reported and in 85 of them the triggering agent is identified: 72 cases are related to flares, 11 to prominence eruptions, 1 to a jet, and 1 to a Moreton Wave.

Flares can excite LAOs in both distant and nearby filaments. In the former case, the flares produce large disturbances that propagate as waves. For example, \citet{gilbert2008} show an event where an X6.5 flare produces a Moreton wave that propagates until it interacts with a filament located more than one solar radius away. This encounter causes the filament to oscillate. 
This interaction has been studied theoretically by \citet{schutgens_numerical_1999,liakh_numerical_2020,Zurbriggen2021} and \citet{liakh_numerical_2023}. 
However, in the majority of cases, LAOs are produced by flares occurring near the filament, like, for example, in the cases described by \citet{jing2003,jing2006} and in many events reported by \citet{luna_gong_2018}. In these cases, there is no obvious disturbance shaking the filament and no wave escaping and propagating through the corona. There are no studies to date that show how flares can excite LAOs without the need for waves to communicate the disturbance.

%%%%%%%%%%%%%%%%%%%%%%%%%%%%%%%%%%%%%%%%%%
%
In this paper, we revisit the event that produced the geoeffective 2015 March 17th, St. Patrick's Day, storm. 
The origin of this storm is an event consisting of a two-step eruption on 14-15 March 2015 \citep{Chandra2017}. The filament is located close to the active region NOAA AR 12297 where weak flares and jets disturb the filament. This activity causes the filament to break into separate segments, each of which supports LAOs. On the 15th there is an eruption of one of the segments which produces a huge oscillation in the %\sout{remnant filament} 
{other segment}. 
%\fminote{[[ '... in the other segment' better than 'in the remnant filament'?]]}
%
%
The series of flares and jets that occur in this active region has been studied to explain the triggers of a coronal mass ejection that takes place in it
\citep[CME;][]{Chandra2017,Bamba2019}. The CME led to the most geoeffective event of solar cycle  24 \citep[Dst=-223 nT,][]{Wang2016,wang_sympathetic_2016,Liu2015,Sindhuja2022} two days after the eruption, on 17 March, St Patrick's Day.
The large oscillation after the CME has been partially studied in two papers. \citet{raes_observations_2017} take this event to show that small phase differences between different parts of the filament can be used to infer properties of a prominence. The periods they obtain are in the range of 59 - 77 minutes. 
With the different periods at different heights, the authors determined that the possible flux rope had the centre of curvature 103 Mm above the solar surface. These values are in agreement with typical prominence cavity heights \citep[see e.g.][]{hudson_stable_1999,gibson_three-dimensional_2010}.
\citet{luna_automatic_2022} also use this event as an example to study the validity of an automatic method for detecting oscillations in filaments. The method works well and determines an oscillation period of $74^{+11}_{-8}$ minutes.
However, these authors did not investigate the mechanisms that produced this huge oscillation.

We aim to provide a comprehensive description of the LAOs occurring on 14 and 15 March and their possible trigger using data from several ground- and space-based observatories. In Sect.~\ref{sec:datadescription} we describe the data used and the methods employed in this work. In Sect.~\ref{sec:description-and-context-of-the-full-event} the context and the phenomena taking place during the two days are given. In Sect.~\ref{sec:oscillations} we analyse the oscillations and jets using the time-distance technique.  Sect.~\ref{sec:triggeringHugeoscillation} describes in detail the triggering of the large oscillation after the partial eruption. In Sect.~\ref{sec:theoretical-analysis} we use a simplified model to give some theoretical estimations of the displacement and velocity amplitudes of the oscillation and an estimation of the flare energy. Finally, in Sect.~\ref{sec:conclusions} the findings are summarised and conclusions are given. 

\section{Description of the telescope data and oscillation analysis methods}\label{sec:datadescription}

\begin{figure*}[!ht]
\centering
\vspace{-2.3cm}\hspace{-0.5cm}\includegraphics[width=\sizefactorfull\textwidth]{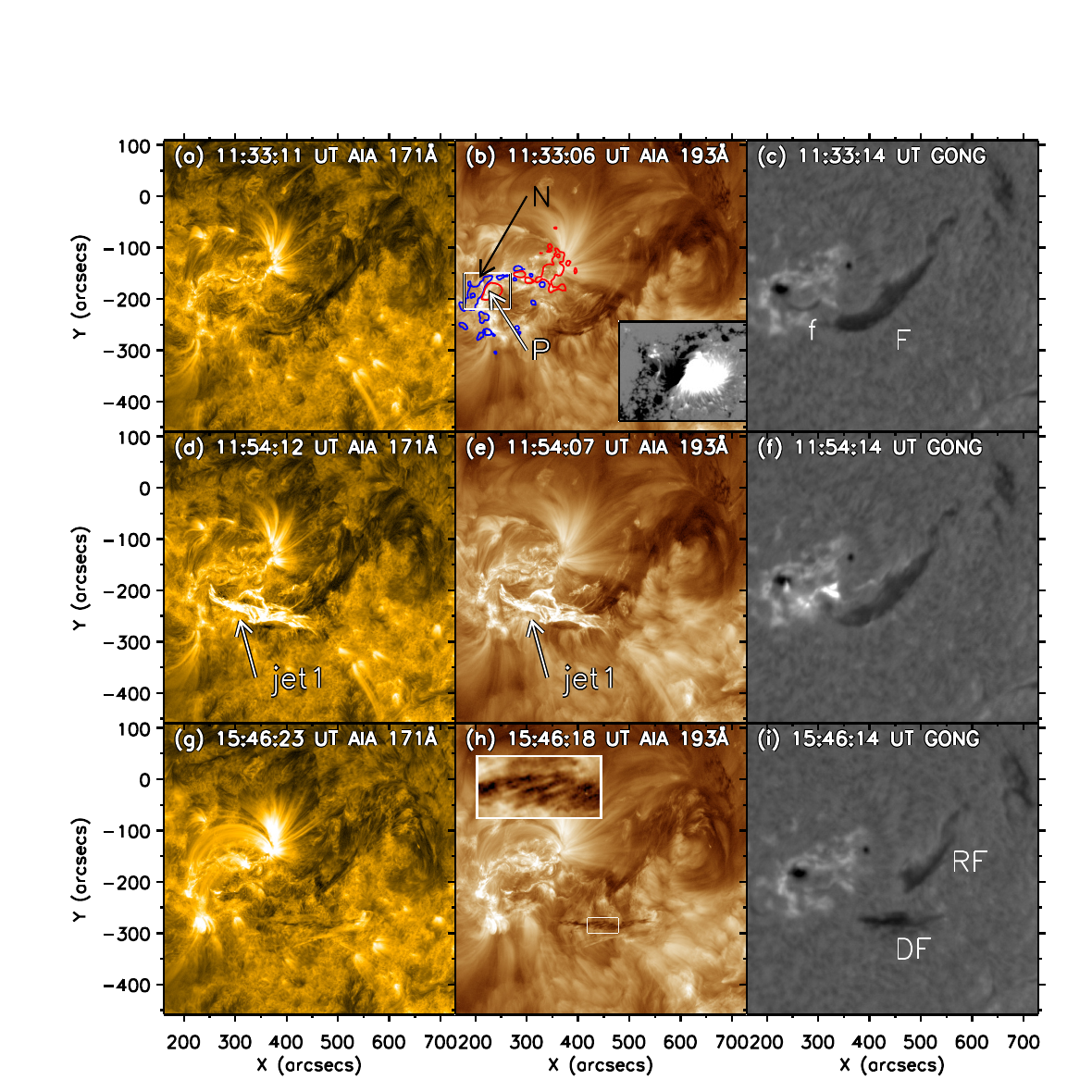}
\caption{Temporal evolution on 14 March 2015 in SDO/AIA 171\AA\ (left column), 193\AA\ (central column) and in GONG \ha (right column).
In panel (b), the images are overlaid with the HMI magnetogram of the AR. 
{In the same panel an inset showing a detail of the HMI magnetogram around the AR central bipole.}
It consists of a bipole with a positive polarity (P, in red) surrounded by negative polarities (N, in dark blue). 
In (c) the image is shown in \ha with a large F filament studied in this work and a small f filament in the AR. The f eruption is related to the first jet on this day.
In (d) and (e) jet1 appears as a bright structure marked with the white arrow. It is visible in the AIA passbands but not in the H$\alpha$, panel (f). 
After the impact of the jet1, the F filament splits into two segments. In the AIA channels ((g) and (h)) both parts can be seen but it is much clearer in H$\alpha$ (i). Both segments are labelled RF and DF. The DF segment detaches from the body of the initial filament and is ejected  up and toward the South. This segment will erupt the next day. In panel (h) inset, a detail of DF is shown. This reveals the structure of the filament in the form of threads oriented along the NW direction.
An animation of this figure is available \href{}{online}.\label{fig:description-day-14}}
\end{figure*}

\begin{figure}[!ht]
	\centering
    \includegraphics[width=\sizefactorhalf\textwidth]{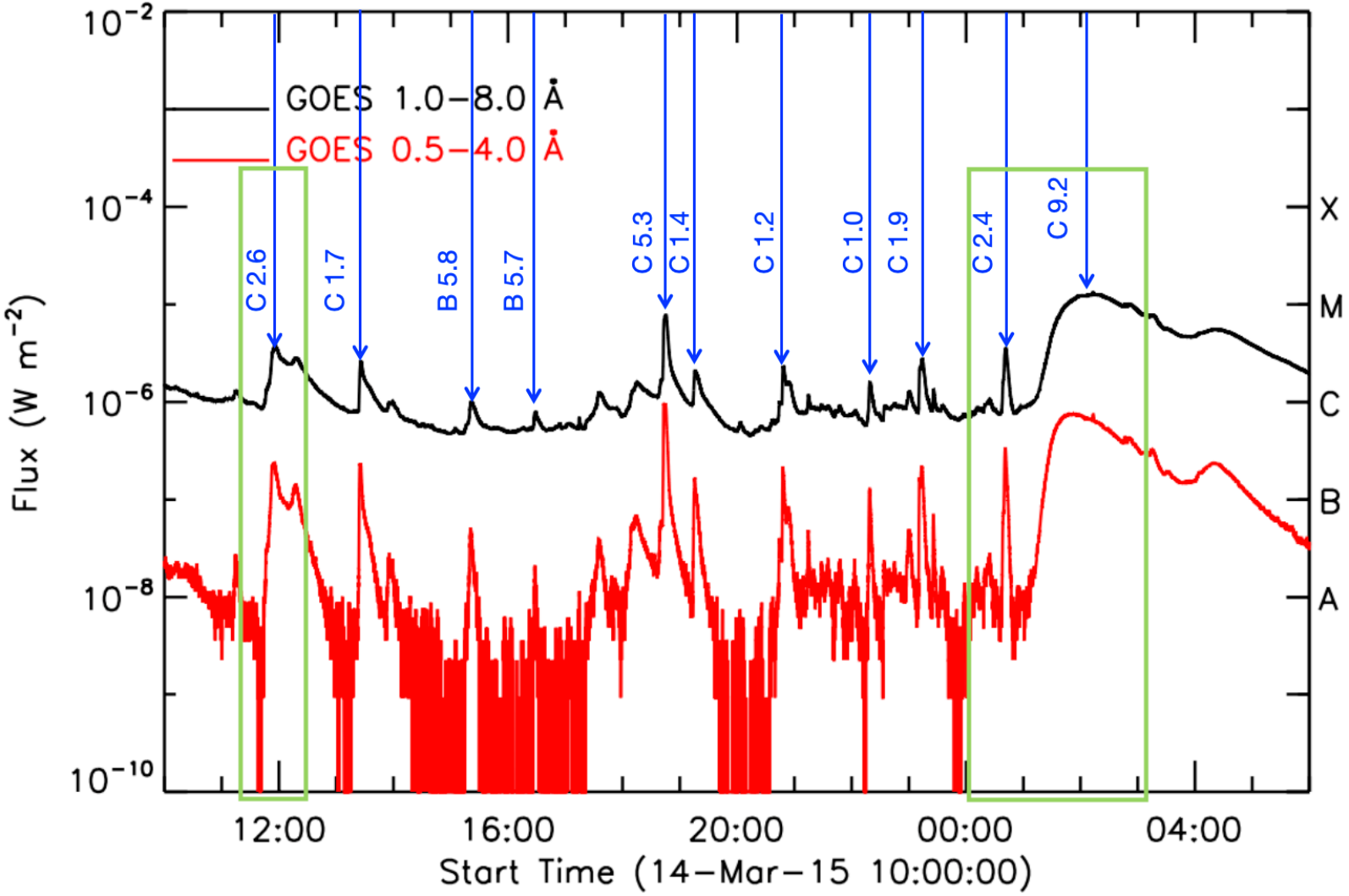}
	\caption{Succession of flares from AR 12297 on March 14-15  2015 registered by GOES. The GOES classification of flares measures the peak flux in the 1--8 \AA\ range (black curve) and in 0.5--4 \AA\ range (red curve). The vertical blue arrows shows the recurring solar flares from AR 12297 for the period March 14 10:00 UT till March 15 6:00 UT. The two green boxes delimit the two flare events we are interested in. The first event contains one C2.6 flare whereas the second includes two flares, of class C2.4 and C9.2, respectively. Jet1 occurs almost simultaneously after the C2.6 flare on day 14 whereas jet2 erupts just after flare C2.4 on day 15.}
	\label{fig:goesdata}
\end{figure}

\begin{figure*}[!ht]
\vspace{-2.3cm}\hspace{-0.5cm}\includegraphics[width=\sizefactorfull\textwidth]{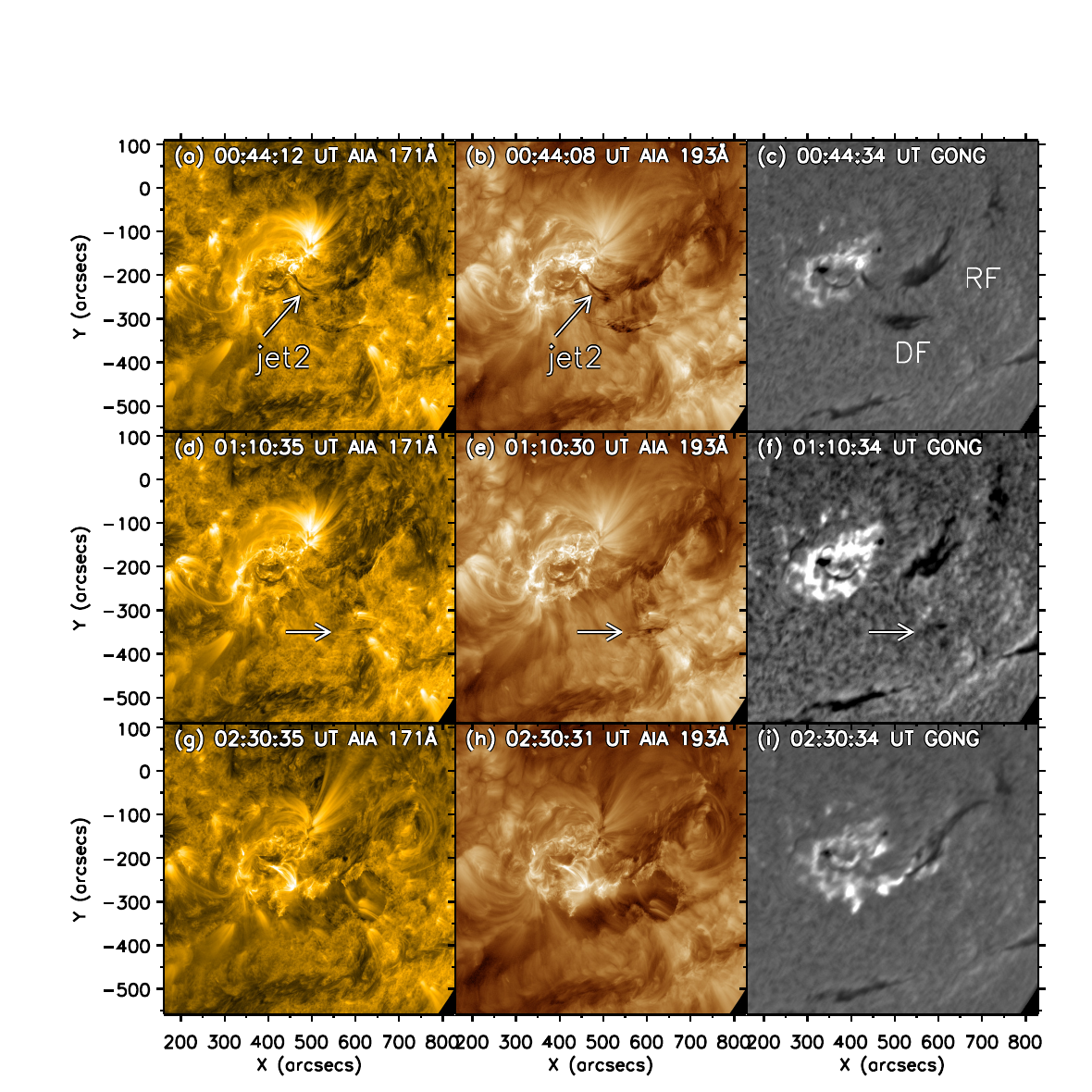}
\caption{Temporal evolution on 15 March 2015 similar to Fig. \ref{fig:description-day-14}. 
Panels (a) and (b) show the jet2 marked by a white arrow. In both AIA panels, the plasma is seen by absorption. A dark structure is also seen in (c). In (c) the two filament segments are seen. Jet2 hits DF and triggers its eruption. In the middle row, DF (white arrow) is slowly rising. In the last column, the fulguration associated with the DF eruption is occurring. In (g) and (h) the postflare loops are visible. In (i) chromospheric brightenings are also seen. These brightenings form two ribbons, one north and one south, although the shape is not well defined. It is clearer in later times and it is recommended to see the animation accompanying this figure. In (i) the RF segment has an elongated shape because it is pushed in a NW direction. 
An animation of this figure is available \href{}{online}.\label{fig:description-day-15}}
\end{figure*}

We use the high spatial and temporal resolution EUV data from the Atmospheric Imaging Assembly instrument \citep[AIA,][]{Lemen2012} on-board the Solar Dynamics Observatory \citep[SDO,][]{Pesnell2012} satellite. AIA  observes the full Sun in seven different UV and EUV wavebands. The {pixel size} and cadence of this instrument are 
0.6~\arcsec\ and 12~s respectively. For better visibility of the filament oscillations, we deconvolved all the AIA images using the AIA point-spread function with the routine \emph{aia$\_$deconvolve$\_$richardsonlucy.pro} available in \emph{Solarsoft} \citep{Freeland1998}. All the AIA images are coaligned after correcting for solar rotation and are mostly plotted in logarithmic intensity scale. 
The deconvolved images are of very good quality and will be the main ones used in this work. However, to improve the visualisation, sometimes the new image enhancement wavelet-optimized whitening \citep[WOW;][]{auchere_image_2023} is applied to some AIA images. We have also explored the multi-scale Gaussian normalization \citep[MGN;][]{Morgan2014} processing technique. MGN tends to equalize the image intensities, however we are interested in highlighting eventual brightenings. On the contrary, WOW accentuates these brightenings and for this reason, we use this technique. In any case, the findings are not affected by the use of this technique because the deconvolved AIA images have very good quality. 

We also use the ground-based  \ha line-centre photospheric data from the National Solar Observatory/Global Oscillation Network Group (NSO/GONG). To cover the Sun for the entire time of the day, there are six network telescopes situated all over the globe (\url{http://gong2.nso.edu}).
The pixel size and temporal resolution of the \ha data are $1$\arcsec\ 
and $1$~min, respectively. \ha observations are the best choice for visualization of the filaments and for tracking their motion; the \ha images show the dark filaments seen in absorption in sharp contrast to the bright chromosphere around them. So, we
use GONG's \ha images to support the AIA EUV observations. 

In addition, we use the longitudinal magnetic field observed with the Helioseismic Magnetic Imager \citep[HMI,][]{Schou2012} onboard SDO to show the evolution of the active region. %\fminote{[[active regions, in plural?]]}
HMI provides the photospheric magnetograms with a pixel size of $0.5\arcsec$ and a cadence of 45 secs. Using the HMI magnetic field contours overplotted on the EUV images, we identify the magnetic environment of the AR and the filament studied. 

%%%%%%%%%%%%%%%%%%%%%%%%%%%%%%%%%%%%%%
To analyze the oscillations and the jets, we apply the time-distance technique as in previous works \citep{luna_observations_2014, Luna_Large-amplitude_2017, luna_gong_2018, joshi_interaction_2023}. 
The time-distance diagrams are constructed with curved artificial slits following the motion of the plasma that will be described in detail in each of the observations \citep[see Appendix at][for further details]{luna_gong_2018}. With these diagrams, we can obtain the oscillation parameters, such as amplitude, damping time, or direction of the motion. The diagrams are constructed using various SDO/AIA passbands mainly with the 171\AA\ filter.
We fit the periodic motions of the time-distance diagram with a damped sinusoidal curve as
\begin{equation}\label{eq:sinusoid}
s(t)= s_0 + A \, e^{-t/\tau} \sin \left( \frac{2 \, \pi}{P} t + \varphi_0 \right) \, ,
\end{equation}
where $s$ is the coordinate along the slit, $A$ is the amplitude, $P$ is the oscillation period, and $\varphi_0$ is the phase at $t=0$.

In this event, we have only detected large amplitude longitudinal oscillations (LALOs){, in which} 
the cold filament plasma moves along the magnetic  field. %
We can apply filament seismology by comparing observations with theoretical models.
\cite{luna_large-amplitude_2012} showed that the period of the longitudinal oscillations is determined by the curvature of the dips that hold the plasma of the prominence \citep[see also][]{Zhang2012} in the so-called pendulum model. 
\citet{luna_extension_2022} extended the pendulum model of \citet{luna_large-amplitude_2012} by incorporating the spatial dependence of the gravity vector {(see Appendix \ref{sectapp2} for a brief description of the model)}. This extension revealed that the original uniform-gravity model is invalid for periods exceeding 60 minutes.
From \cite{luna_extension_2022} we can write the relation between the period $P$ and the curvature of the dips $R$, as {(see Eq. \eqref{eq:new-pendulum-appendix})}
\begin{equation}\label{eq:new-pendulum}
    \frac{R}{R_\odot}=\frac{\left(P/P_\odot\right)^2}{1-\left(P/P_\odot\right)^2} \, ,
\end{equation}
where $R_\odot$ is the solar radius and $P_\odot =2\pi\sqrt{\rsun/g_0}=$ 167 minutes {(see Eq. \eqref{eq:period-sun})}. In addition, the authors discovered that the periods of the longitudinal oscillations cannot exceed $P_\odot$.
The authors also determined a minimum value of the magnetic field of the dipped lines in order to have magnetic support of the cool mass. Using {Eq. \eqref{eq:magnetic-field-seismology-appendix}} we obtain
\begin{equation}\label{eq:seismology-magnetic-field}
    B (\mathrm{Gauss}) = \left(47\pm 24\right) \frac{P/P_\odot}{\sqrt{1-\left(P/P_\odot\right)^2}} \, .
\end{equation}
The numerical values in brackets are computed assuming a typical density range for prominences of $\rho=2 \times 10^{-11} - 2 \times 10^{-10} \mathrm{~ kg ~ m^{-3}}$. We consider the prominence density to be the most important source of uncertainty in the estimated magnetic field because the uncertainties associated with the fit are smaller than those related to the above density range. In the following, we use both equations to compute the radius of curvature and magnetic field strength.

\section{Active region and Filament on March 14 and 15}\label{sec:description-and-context-of-the-full-event}
The filament activation and the following eruption are produced in two steps or phases, the first one on March 14th and the second one on the 15th \citep{Chandra2017}.  In the following, we describe the phenomena occurring in either phase separately. 
This description provides the context in which the oscillation occurs. The oscillations in the two phases are described in Sect. \ref{sec:oscillations}.

\subsection{First Phase on March 14th}\label{subsec:firstphase-context}

Figure \ref{fig:description-day-14} shows the temporal evolution of the filament on March 14th in three wavebands AIA 171\AA, 193\AA\ and GONG \ha. In panel \ref{fig:description-day-14}(b), the HMI magnetogram overlay shows the magnetic configuration of the event. 
%\fminote{'The HMI magnetogram overlay' is confusing: it sounds like the little inset, which is also an HMI magnetogram. Suggestion: 
{In panel \ref{fig:description-day-14}(b), isolines from the HMI magnetogram are overlaid (in blue and red for the negative and positive polarities respectively) to mark the magnetic configuration of the event. This area corresponds to the huge}
and complex AR NOAA 12297 situated on the solar disk (S17W26) with leading positive polarities and a large following sunspot in a $\delta$ magnetic configuration.
The AR centre consists {(see inset in Fig. \ref{fig:description-day-14}(b))} of a bipole with positive polarity (P) surrounded by negative polarities (N) resulting from the emergence of new flux inside a remnant AR; this constitutes a very favourable configuration for flare and jet activity \citep[see further details in][]{wang_sympathetic_2016,Chandra2017}.
In the neighbourhood of the AR, an extended filament (labelled 'F' in Fig. \ref{fig:description-day-14}(c)) lay pointing to the main sunspot and 
%\sout{reaching an elongation of up to several degrees longitude westward; it extended}
extending westward over more than 400\arcsec\ in the FOV, as shown in Fig.~\ref{fig:description-day-14}(c). The filament is intermediate and it is oriented in the North-West direction; one of its endpoints is near the AR; the other shows a turn to the south.

The filament is very wide in the GONG H$\alpha$ images in the southern hemisphere and also seems to be of sinistral magnetic class due to the appearance of the barbs (regularly spaced feet of the filament) towards the left. As the GONG filter has a large bandpass, this suggests that the filament can be seen in different wavelengths in the H$\alpha$ wings indicating perturbations with blue and redshifts, enlarging the line profiles  \citep{Schmieder1985}.
The images in 171\AA\ and 193\AA\ (Figs. \ref{fig:description-day-14}(a) and \ref{fig:description-day-14}(b)) show the absorbing filament and a larger dimmed band forming the sigmoid-shaped filament channel.
In addition to the large filament, there is also a small narrow  AR filament close to the positive spot P labelled with f in Fig.~\ref{fig:description-day-14}(c). This small filament is related to the first jet that we will describe in the following.

This AR showed a lot of activity with B- and C-class flares as shown by the GOES data given in Fig~\ref{fig:goesdata}. 
The relevant events for the current study are a C2.6 flare taking place at 11:44 UT (green box on the left in the figure), and C2.4 and C9.2 flares on March 15 (green box on the right) described in Sect. \ref{subsubsec:context-second-phase}. This activity corresponds to the first and second phases of the event.

Consecutively after the small C2.6 flare at 11:44 UT, there is the ejection of a jet  hereafter referred to as jet1. This jet is clearly visible in Figs. \ref{fig:description-day-14}(d) and \ref{fig:description-day-14}(e). Both images show the well-developed jet 10 minutes after ejection. Jet1 appears to be launched from the sunspot's penumbra.
This jet moves to the South-West and some threads are detached and reconnected with field lines of the filaments (Fig. \ref{fig:description-day-14}(d) and (e)). In Sect. \ref{sec:oscillations} the jet velocity  is estimated to be $\speedjetone$ using the time-distance technique.   
This activity corresponds to the double peak in Fig. \ref{fig:goesdata} around 12:00 UT marked with the green box on the left. 
The \ha images show a bright sigmoid which is the signature of a flare (Fig. \ref{fig:description-day-14}(f)).
The flare that will eventually produce jet1 could be related to the eruption of the small filament labelled f in panel (c) that has disappeared in panel (f). This eruption produced a slow CME (350 $\kms$) as observed by the LASCO coronagraph \citep{Liu2015,wang_sympathetic_2016}. 

The interaction of jet1 and filament F results in the detachment of a large portion of the filament from its main body (see Fig. \ref{fig:description-day-14}(g)-(i)). Henceforth we will call the two resulting pieces {\it the remaining filament} (RF) and {\it the detached filament} \citep[DF; called $F_{1}$ in][]{Chandra2017} as marked on panel~\ref{fig:description-day-14}(i). The DF can also be seen in 171\AA\ and 193\AA\ where its thread-like structure can be discerned. This DF segment changes its position until it reaches a new equilibrium. It will remain in more or less the same position until the next day when another jet triggers its eruption.

\subsection{Second  Phase on March 15th}\label{subsubsec:context-second-phase}

Figure \ref{fig:description-day-15} shows the temporal evolution of the second phase, which took place on March 15th. An eruption occurred that is considered to be the trigger for the great St Patrick's Day geomagnetic storm on the 17th (see Sect. \ref{sec:introduction}). For a full and detailed description of the mechanisms that trigger this eruption see \cite{Bamba2019}.
Fig. \ref{fig:goesdata} (green rectangle on the right) shows two peaks, one of them associated with a C2.4 flare at 00:34 UT and the other corresponding to a long-duration C9.2 flare starting at 01:15 UT and lasting until 05:00 UT with a secondary peak around 04:20 UT. 
Simultaneously with the C2.6 flare at 00:34 UT a small filament or jet (hereafter referred to as jet2) escapes and hits the large filament in the South of the region. 
We see the absorbing plasma and parallel bright strands in the AIA 171, 193\AA\ filters (Figs. \ref{fig:description-day-15}(a) and \ref{fig:description-day-15}(b)) and also cool material in \ha (Fig. \ref{fig:description-day-15}(c)) suggesting that it is a surge, with a small filament at the base. 
Jet2 appears to impact RF and also precipitate the eruption of DF.
Around 00:45 UT, DF starts to rise with increasing speed. In Fig.~\ref{fig:description-day-15}, it can be seen that DF has moved and is slightly more separated from RF.
In panels \ref{fig:description-day-15}(d) to \ref{fig:description-day-15}(f), we have marked the position of DF with a white arrow in the three wavebands. In 193\AA, DF is clearly visible, however in 171\AA\ and \ha it is more difficult to be identified. In \ha we have increased the contrast of the image for better visualisation. 
It can be seen that DF has already started to rise slowly at first but considerably faster shortly afterwards. \cite{Chandra2017} made a detailed analysis of the DF eruption and estimated that the projected rise velocity is 70 $\kms$ in the last phase of the eruption.

The C9.1 flare starts at around 1:10 UT in the AR and two ribbons extend East West. In panels \ref{fig:description-day-15}(d)-\ref{fig:description-day-15}(f), the bright structures are visible but, in the following panels \ref{fig:description-day-15}(g)-\ref{fig:description-day-15}(i) the ribbons are clearer.
Brightening is also observed in 171\AA\ in the whole region around the AR and also in the RF filament (as we will see in Fig. \ref{fig:multiplanel171-20150315}). This will be considered in detail in Sect. \ref{sec:triggeringHugeoscillation}. When the flare reaches its maximum at 2:00 UT (see Fig. \ref{fig:goesdata}) postflare loops appear and are seen in panels \ref{fig:description-day-15}(g) and \ref{fig:description-day-15}(h).

 \section{Time-distance analysis}\label{sec:oscillations} 
In this section we study the oscillations and jets that occur on the 14th and 15th of March. Our aim is to qualify and quantify the characteristics of the oscillations and put them in the context of the activity of the region. 

\begin{table*}
\caption{Best-fit and seismology parameters of the three oscillations reported in this work. The first column shows the day and the segment where the oscillation occurs. The second shows which phenomenon may be triggering the oscillation. Columns 3 to 5 show the parameters of the oscillation fit, period, damping time and velocity amplitude respectively and their uncertainty. Columns 6 and 7 show the results of applying seismology, the radius of curvature and the magnetic field intensity respectively as explained in Sect. \ref{sec:datadescription}.}              % title of Table
\label{table:1}      % is used to refer this table in the text
\centering                                      % used for centering table
\begin{tabular}{c c c c c c c}          % centered columns (4 columns)
\hline\hline                        % inserts double horizontal lines
Event & Trigger & $P(\mins)$ & $\tau(\mins)$ & $V(\kms)$ & $R(\Mm)$ & $B(\G)$\\    % table heading
\hline                                   % inserts single horizontal line
    March 14, DF & DF position change & $\periodlalofirstphaseDFnu$ & $\dampingtimelalofirstphaseDFnu$ & $\velocitylalofirstphaseDFnu$ & $\radiuslalofirstphaseDFnu$ & $\magneticlalofirstphaseDFnu$\\      % inserting body of the table
    March 14, RF & Jet1 & $\periodlalofirstphaseRFnu$      & $\dampingtimelalofirstphaseRFnu$ & $\velocitylalofirstphaseRFnu$ & $\radiuslalofirstphaseRFnu$& $\magneticlalofirstphaseRFnu$\\
    March 15, RF & Flare & $\periodlalosecondphasenu$        & $\dampingtimelalosecondphasenu$ & $\velocitylalosecondphasenu$ & $\radiuslalosecondphasenu$ & $\magneticlalosecondphasenu$\\
\hline                                             %inserts single line
\end{tabular}
\end{table*}

\begin{figure}[!ht]
	\centering
	\includegraphics
	[width=\sizefactorhalf\textwidth]
	{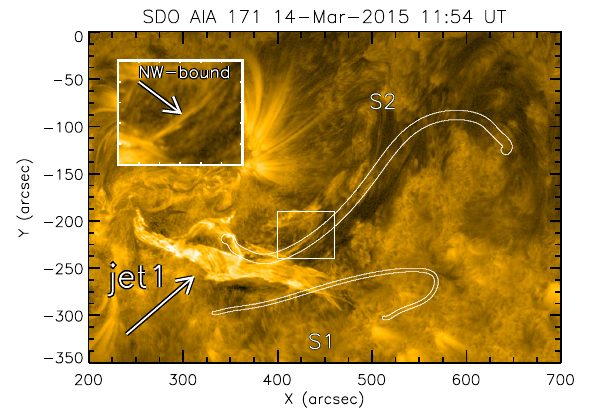}
	\caption{AIA 171\AA\ image showing the artificial slits S1 and S2 used to study the oscillations on DF and RF respectively. S1 has a width of 4 arcsecs whereas S2 has a width of 10 arcsecs. The figure shows jet1 (white arrow) interacting with the filament. This jet will split the filament in two, DF and RF (see Fig. \ref{fig:description-day-14}). However, this shows an early stage and DF has not yet moved to its new position under the S1 slit. Part of the jet1 flows are directed in a NW direction along S2. The inset shows the white box area enlarged to show the details of the NW-bound jet. 
	}
	\label{fig:mapwithmask-firstevent}
\end{figure}
\begin{figure*}[!h]
	\centering
	\includegraphics
	[width=\sizefactorfull\textwidth]
	{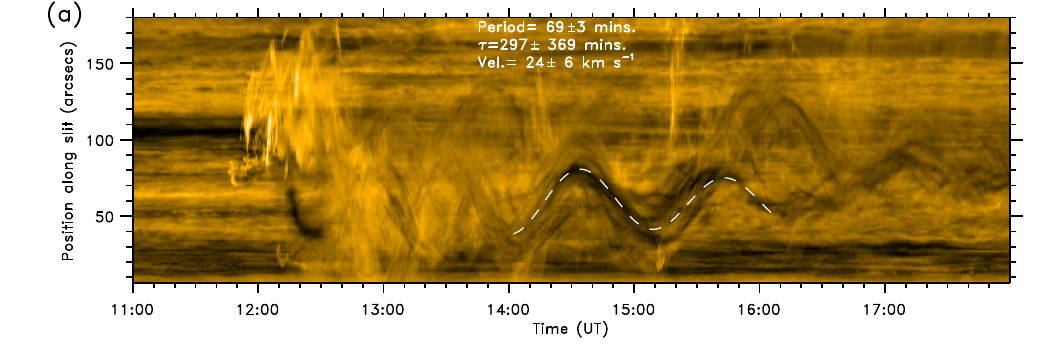}
	\includegraphics
	[width=\sizefactorfull\textwidth]{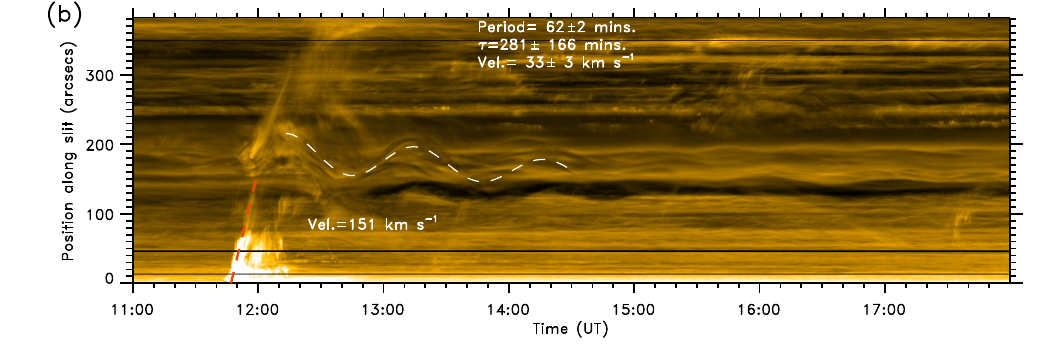}
	\caption{Time-distance diagrams of March 14th constructed with the artificial slits shown in Fig. \ref{fig:mapwithmask-firstevent}. In (a) the diagram for S1 over DF is shown and in (b) the diagram in S2 over RF both in AIA 171\AA. In this passband the prominence plasma appears in absorption as dark bands. The white dashed line shows the oscillation fit using Eq. \eqref{eq:sinusoid}. The best fit parameters of the oscillation are shown over both panels. In (b) the fit is also shown by plotting jet1 with a red dashed line. This gives the jet propagation velocity shown on the panel.}
	\label{fig:time-distance-oscillation-march14}
\end{figure*}
\subsection{First phase}\label{subsec:oscillationsofthefirtsphase}
The triggering is related to the C2.6 flare at 11:44UT on March 14th, the first jet (jet1) and the detachment of part of the filament (DF).  
As we have described in Sect. \ref{sec:description-and-context-of-the-full-event}, jet1 splits the large F-filament into two parts, labelled RF and DF. The DF segment changes its position until it reaches a new equilibrium \citep{Chandra2017}. 
This results in oscillations in both segments. The movement of DF to a new position probably shakes the plasma of the prominence and contributes to the oscillation. 

To study the oscillations of RF and DF we use the time-distance technique with the 171\AA\ data. Fig. \ref{fig:mapwithmask-firstevent} shows one frame of the AIA 171\AA\ data where jet1 appears as a bright structure. Overplotted on the image are the white contours of the artificial slits used to extract the intensity as a function of time. 
To define the curved slit in the images, we tracked the path of the oscillations by visual inspection. We trace the oscillation motion of the filament segment we are interested in and also include the trajectory of the jet. 
With the artificial slits S1 and S2, we track the oscillations on DF and RF respectively. Note that DF is not yet under S1 at the instant shown in the figure. However, after the jet, DF moves to a new position under S1, and we can follow its oscillation. 
It is unclear how jet1 triggers the oscillation in DF as the jet does not appear to flow along the field lines where DF is located. We think that the structural changes that occur after the jet1 ejection cause the oscillation of the cold plasma in DF. 
As we see in Fig. \ref{fig:mapwithmask-firstevent} most of the jet flows are oriented in the SW direction and some are moving to the NW  roughly following the direction of S2. The NW-bound flows move along S2 until they interact with the RF segment.
To get a better view of the NW-bound flows, we have enlarged the area marked in the white box and shown it in the inset of the same Fig. \ref{fig:mapwithmask-firstevent}: the bright elongated structure in 171\AA\ is clearly visible in it. In addition, part of the jet1 flows move towards RF, producing its oscillation.

Fig. \ref{fig:time-distance-oscillation-march14} shows the resulting time-distance diagrams for both the S1 and S2 slits.
In diagram \ref{fig:time-distance-oscillation-march14}(a) we see that the jet1 perturbation reaches S1 around 12:00 UT as a collection of bright features. Between 13:00 and 14:00 UT, we can discern the apparently oscillatory motion of dark features. They are out of phase and it is difficult to adjust any clear oscillation to them. However, from 14:00 UT  onwards the oscillatory motion is clear and can be seen until after 16.00 UT. In that time range, we can follow the oscillation and fit a damped sinusoidal curve using Eq. \eqref{eq:sinusoid}. 
Table \ref{table:1} shows all the best-fit parameters of Eq (\ref{eq:sinusoid}) for all the oscillations studied in this work.
In Sect. \ref{subsec:seismology} the results obtained with prominence seismology are presented and discussed.
In Fig. \ref{fig:description-day-14}(h) an inset is plotted showing the enlarged area delimited by the white box over the spine of the filament. In this inset, we observe elongated structures probably aligned with the magnetic field, forming an angle with the spine of the filament. These appear to be individual cool threads, or clusters thereof, seen in absorption.
In the observed oscillations the threads move along their axis, so we can assume that the oscillations are longitudinal. The perturbation probably also induces transverse oscillations but we have not been able to detect them. 
Correspondingly, Fig.~\ref{fig:time-distance-oscillation-march14}(b) shows the time-distance diagram of S2. In this case, we can identify the triggering, which is the NW-bound flows of jet1. 
The interaction resembles the one described in \citet{joshi_interaction_2023} which fits the \citet{Luna2021} model for the jet-prominence interaction. With a single slit, we can track the jet and the oscillation. 
The jet starts at 11:44 UT and in the diagram it appears as a collection of straight bright features. These have  an inclination that depends on the propagation speed. 
We determine the jet velocity by fitting a straight line (red dashed line) over these brightenings. We conclude that the jet1 flows move along S2 with a velocity of $\speedjetone$. From the diagram we also appreciate that jet1 reaches the filament at the slit position $s=180$ arcsec a few minutes after the jet1 ejection, thereby triggering the oscillation. Part of the hot jet flows continue propagating along S2 as seen in the bright straight lines above $s=180$ arcsecs in the diagram.
Between 12:00 and 13:30 UT we can clearly identify the oscillation. In contrast, from 13:30 to 14:30 UT, it is less clear and after 14:30 UT no dark thread oscillation is visible. We have fitted a damped sinusoidal as in Equation~\ref{eq:sinusoid} to the oscillatory pattern (white dashed line) obtaining the parameter values shown in Table \ref{table:1}.

\subsection{Second phase}\label{subsec:oscillationsofthesecondphase}
\begin{figure}[!ht]
	\centering\vspace{-2.2cm}
	\includegraphics
	[width=\sizefactorhalf\textwidth]
	{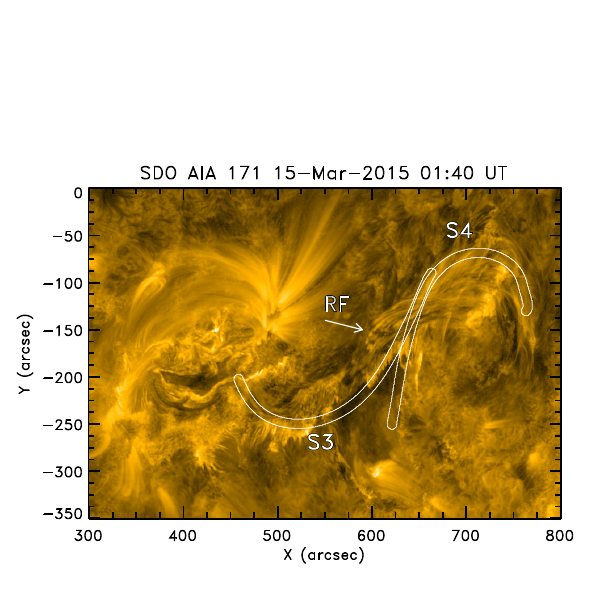}
	\caption{AIA 171 \AA\ image showing the filament on 15 March similarly to Fig. \ref{fig:mapwithmask-firstevent}. The artificial slits S3 and S4 are shown. S3 is used to study the propagation of the jet2 while S4 is used to study the oscillations of the RF segment. Both slits have a width of 10 arcsecs.}
	\label{fig:mapwithmask-secondevent}
\end{figure}

In this section, we study the characteristics of the oscillations which occur the next day. The trigger of the oscillations will be analysed in a separate section  (Section 5).
On March 15th the DF segment erupts 
at 01:15 UT. This eruption produces a huge LALO in the RF segment. The same eruption is also responsible for the St. Patrick's Day geomagnetic storm on Earth on March 17th.
Figure \ref{fig:mapwithmask-secondevent} shows the RF filament during the triggering phase.
Bright loops outlining the structure of the filament channel are visible.
Dark strands are seen along with the bright plasma. The dark strands will be part of the prominence that is displaced towards the northwest end of the filament channel.
In the figure, we have drawn the two slits used in the second phase. S3 was used to study the jet produced in the AR and S4 was used to track the oscillation.

\begin{figure*}[!ht]
	\includegraphics
	[width=\sizefactorfull\textwidth]
	{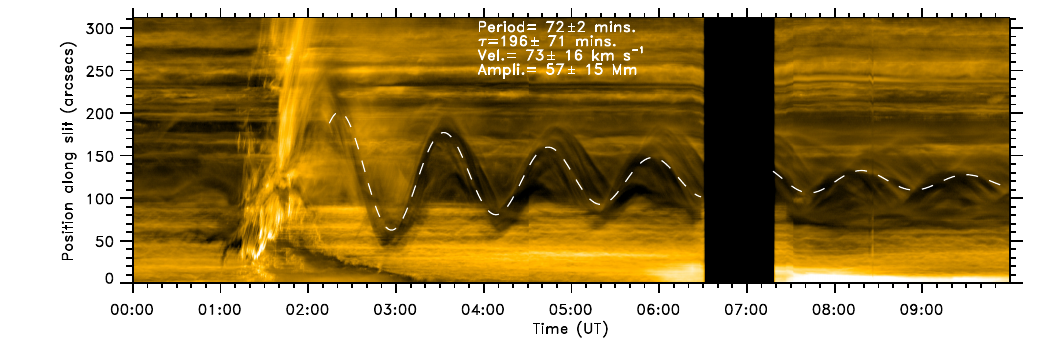}
	\caption{Similar to Fig. \ref{fig:time-distance-oscillation-march14} the time-distance diagram in 171\AA\ along slit S4 is shown. The oscillation best-fit parameters are shown in the figure with the white dashed line. Bright structures are seen between 01:10 and 02:10 UT which is the impulsive phase when the oscillation is triggered. After 2:10 UT the absorbing plasma in the filament moves periodically around y=120 arcsecs. Between 6:30 and 7:20 UT, there is a gap with no SDO data.}
	\label{fig:timedistance-march15}
\end{figure*}

Figure \ref{fig:timedistance-march15} shows the time-distance diagram constructed with S4. This diagram shows that the triggering of the oscillation occurs approximately between 01:10 and 02:10 UT. It is interesting to note that the impulse occurs over quite a long time and is not an instantaneous process in contrast with the case produced by jet 1 shown in Fig. \ref{fig:time-distance-oscillation-march14}(b). After 02:10 UT the oscillation of the cold plasma that appears in absorption is already visible. The oscillation is very clear and extends beyond the time interval considered in the figure. It can be seen that several threads move in phase during the first cycles of the oscillation but end up losing global coherence at the end of the diagram. We have used the technique discussed in Sect. \ref{sec:datadescription} to track the movement of the filament. Fitting this data with Eq. (\ref{eq:sinusoid}) we obtain the curve shown as a white dashed line. The best-fit parameters of the oscillation are given in Table \ref{table:1}.
The amplitude of the velocity is quite large and considering that the whole prominence is involved in the oscillation the energy must be enormous.
In Sect. \ref{sec:theoretical-analysis}, considerations regarding the energy associated with these oscillations are presented.

\subsection{Seismological Analysis}\label{subsec:seismology}
We apply a seismological analysis of the studied oscillations as described in Sect. \ref{sec:datadescription}.  
With the oscillation periods (see Table \ref{table:1}), we obtain information on the radius of curvature of the magnetic dips, $R$, as well as the value of the field strength, $B$.
The periods {are similar} for the three cases and range from  62 to 72 minutes. 
With Eq. (\ref{eq:new-pendulum}) we {then} obtain radii of curvature {for} the three events {which range} from $111$ to $159$~Mm. We see that 
%\sout{with}
the RF oscillations {of March 14th} 
%\sout{on day 14, we obtain} 
{yield} a radius of $111$~Mm. By the next day, the radius of curvature {of RF} {has increased} 
%\sout{increases}
to $159$~Mm.  The magnetic field {strength has} also {increased
a little}, namely  
%\sout{in intensity}
from $19$ to $22$~Gauss. 
{The reason for the changes may be that we are analysing different field lines on each day; another}
%\sout{We will probably be analysing different field lines on both days, getting different results. Another}
option is that this difference may be associated with the structural changes that occur in the filament during the eruption on {March 15th}.
However, further study of the oscillations would be necessary to confirm {which option is most likely to be correct.}
% that they are associated with structural changes during the eruption. %\fminote{ ... necessary to confirm which option is most likely to be correct ?}
The radius of curvature obtained for the DF segment is $143$~Mm {(Table~\ref{table:1}, first line)}. Interestingly, this portion that has partially erupted has a  radius of curvature {similar to that of} the remaining part. This may indicate that the dips of this detached segment are not changing considerably during this process.

\section{The trigger for the March 15th huge oscillation}\label{sec:triggeringHugeoscillation}
\begin{figure}[!ht]
	\centering
	\includegraphics
	[width=\sizefactorhalf\textwidth]
{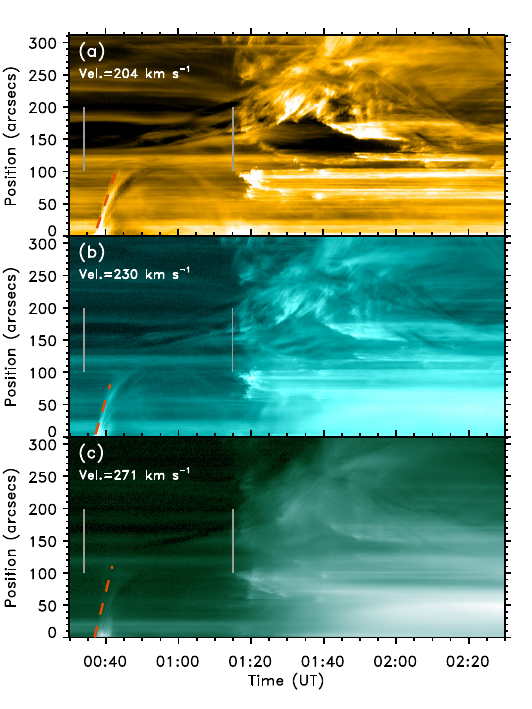}
	\caption{Time-distance diagram for jet2 along {the} artificial slit S3 (Fig. \ref{fig:mapwithmask-secondevent}) in 171\AA\ (panel a), 131\AA\ (panel b), and 94\AA\ (panel c). In all the panels, the two vertical grey lines mark the times of the first C2.4 flare at 00:34UT and the second C9.2 flare starting at 01:15UT. Also in the three panels, fitting a straight line (red dashed line) we estimate the speed of the jet2 flows on each passband. The jet2 propagation velocities are shown on each panel. The average value for all three passbands is $\speedjettwo$.\label{fig:timedistance-march15-jet-possible-triggering}}
\end{figure}
The main motivation for this work is to understand the triggering of the huge oscillation on March 15th. 
In this section, we show in detail the observational evidence that can explain how the filament {is set into motion.}
%\sout{starts to oscillate.}
%\fminote{alternative: '... explain how the filament is set into motion' }
%
As discussed in Sects. \ref{subsubsec:context-second-phase} and \ref{subsec:oscillationsofthesecondphase} the oscillation occurs following the C2.4 flare and jet2 at 00:34  UT. 

Fig. \ref{fig:timedistance-march15-jet-possible-triggering} shows the time-distance diagram for S3 following the jet2 flows in 171\AA, 131\AA\ and 94\AA. 
There are two vertical grey lines representing the times when the first flare C2.4 (left) occurs and the second flare C9.2 (right) starts. 
We measure the propagation speed of jet2 in the three passbands.
Around 00:34 UT the jet starts to propagate along S3 with a speed above $200 \kms$ in all the channels. As we can see, the measured velocity depends on the AIA channel.
However, this does not necessarily imply a dependence of speed on temperature, as the method of measuring inclination is itself quite uncertain. We can give an estimate of the jet speed as the average $\speedjettwo$ of the three channels.
At 00:50 UT the flow becomes quite diffuse and part of the ejected plasma falls back down again (Fig.~\ref{fig:timedistance-march15-jet-possible-triggering}(a)). This is a parabolic movement which can be clearly seen in the 171\AA\ and 131\AA\ diagrams but not in the 94\AA\ one.
At around 01:20 UT, bright structures appear in the filament, like those shown in Fig. \ref{fig:timedistance-march15}. These brightenings are very clear in 171\AA\ and 131\AA\ while in 94\AA\ they are more diffuse. These brightenings are associated with the impulsive phase {in which} %\sout{where} 
the oscillation is triggered. It can be seen that jet2 and the triggering are well separated in time, in contrast %\sout{with} 
{to} the effect of jet1 shown in Fig \ref{fig:time-distance-oscillation-march14}(b). 
The diagram also shows the filament as a dark horizontal band centered at y=150 arcsecs before 01:20 UT.
It is difficult to determine the exact moment when jet2 reaches the filament.
However, we can estimate that it could be between 00:45 and 01:00 UT. The first time corresponds to the intersection of the prolongation of the red dashed curve with the black filament at y=150 arcsecs. The second is the time {when the jet visually} 
%\sout{that visually the jet} 
reaches RF as we will show later.
This indicates that there is a delay {of some $20$ minutes} between the arrival of the jet2 flows and the impulsive phase. 
%\sout{of around 20 minutes.}
%
As discussed in Sect. \ref{subsubsec:context-second-phase}, between 00:34 UT and the impulsive phase of the oscillation 
the DF segment erupts and the C9.2 flare occurs (right vertical grey line in Fig. \ref{fig:timedistance-march15-jet-possible-triggering}), starting at 1:15 UT and peaking at around 02:00 UT. This coincides with the impulsive phase of the oscillation.
With this figure, it is clear that %\sout{the} 
jet2 is related to the first flare but it is when the second flare starts that the brightenings on the filament are visible.
In channels 131\AA\ and 94\AA, Figs. \ref{fig:timedistance-march15-jet-possible-triggering}(b) and \ref{fig:timedistance-march15-jet-possible-triggering}(c) respectively, no features are seen flowing along the slit. However, the plasma below y=150 arcsecs (i.e. to the South-East of the filament) has strong emission after the second flare. This suggests a heating of this area.
In all the panels, between the two vertical lines (between the two flares) the filament already starts to move slowly, i.e., the dark band moves along S3. This could reflect that the whole structure starts to change during the DR eruption.

\begin{figure*}[!ht]
	%\vspace{-1.cm}
    \centering
	\includegraphics
	[width=\sizefactorfull\textwidth]
	{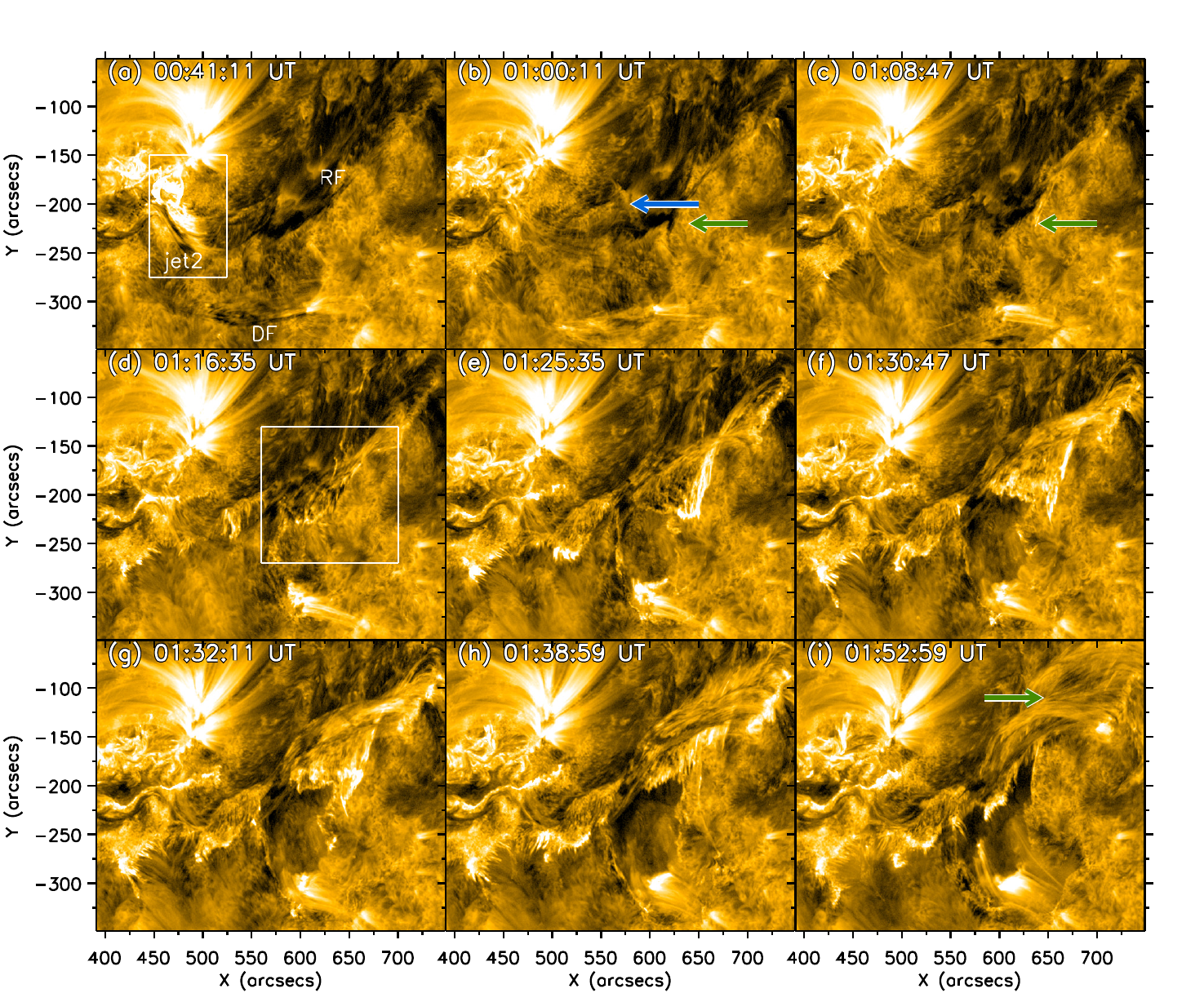}
	\caption{Temporal evolution of the eruption of DF and the triggering of the oscillations in the RF segment in 171\AA\ data.
In (a) the absorbing plasma of the RF and DF segments shown in the figure is visible. In the same panel, the white box delimits the area where the jet2 is located.
In panel (b) DF jet2 has already impacted DF and this segment is already erupting. Jet2 also impacts RF producing a slight brightening in a small area on the RF spine (blue arrow). During the eruption of DF, it seems that changes also occur in RF, such as the appearance of a protrusion on its west side (green arrow).
This same protrusion (green arrow) becomes bright shortly afterwards.
In (d), within the white box, many brightening on RF can be seen.
In the next panels from (e) to (h) the brightenings are massive over the whole structure and in (i) loops can be seen in the NW area (green arrow) outlining a kind of flux rope structure.
An animation of this figure is available \href{}{online}.} \label{fig:multiplanel171-20150315}
\end{figure*}

Fig. \ref{fig:multiplanel171-20150315} shows the temporal evolution in the AIA 171\AA\ passband. In panel \ref{fig:multiplanel171-20150315}(a), we can distinguish both segments RF and DF; %\sout{The} 
jet2 can also be identified on the left of the {panel} 
%\sout{image} 
and is marked with a white box. We can see that it is composed of bright hot plasma but also of dark plasma seen in absorption.
These flows continue propagating forming arched curves in the direction of RF. As we have mentioned in the previous paragraph part of the plasma moves back to the origin of the jet but part also interacts with the filament. In panel \ref{fig:multiplanel171-20150315}(b), we have marked with a blue arrow a bright region over RF which is the region where the flows reach the prominence. There is no evidence of  oscillation triggering in that region directly following the impact of the jet.
After jet2, DF appears to start to rise slowly and roll/unroll around itself as reported by \citet{wang_sympathetic_2016}. 
It seems that the structure of the RF prominence changes during DF's rise (see Animation of Fig. \ref{fig:multiplanel171-20150315}). The cold plasma can be seen to move slowly. A protrusion marked with a green arrow also appears in \ref{fig:multiplanel171-20150315}(b). %
A few minutes later this protrusion becomes
bright, indicating that some heating is occurring in it; in fact, this seems to be  the first region on RF where the triggering starts (see panel \ref{fig:multiplanel171-20150315}(c); green arrow), and it is worth remarking that
this region is relatively far from the impact of jet2 on RF (blue arrow on Fig. \ref{fig:multiplanel171-20150315}(b)). 
In the next panel, (d), the brightening starts to appear in the south of the RF, as marked with a white square. In (e) the brightenings are widespread and in (f) they are massive {massively apparent} in the entire RF. 
In these images, the bright structures can be seen to have a fibrillar structure; these fibrils are probably the filament threads, which have become bright by some mechanism during this phase. We will come back to these brightenings at the end of this section.
In the following panels \ref{fig:multiplanel171-20150315}(g)-\ref{fig:multiplanel171-20150315}(i) the bright structures can still be seen delineating the twisted structure of the magnetic field in the filament channel.
This is especially clear in panel \ref{fig:multiplanel171-20150315}(i) marked with a green arrow. 
This sequence of images clearly shows that after the fading of the jet2 flows the triggering continues for more than an hour. This excludes jet2 as the direct cause of the oscillation
Still, jet2 seems to be anyway involved, {possibly}  %either 
%\fminote{'either' does not match the sentence (either ...  or ... ) and must be changed. I assume you mean 'possibly', or 'probably', or 'perhaps' ?} 
by activating other processes that ultimately trigger the oscillation.

\begin{figure*}[!ht]
	\vspace{-3cm}\hspace{-0.5cm}\centering
	\includegraphics
	[width=\sizefactorfull\textwidth]	{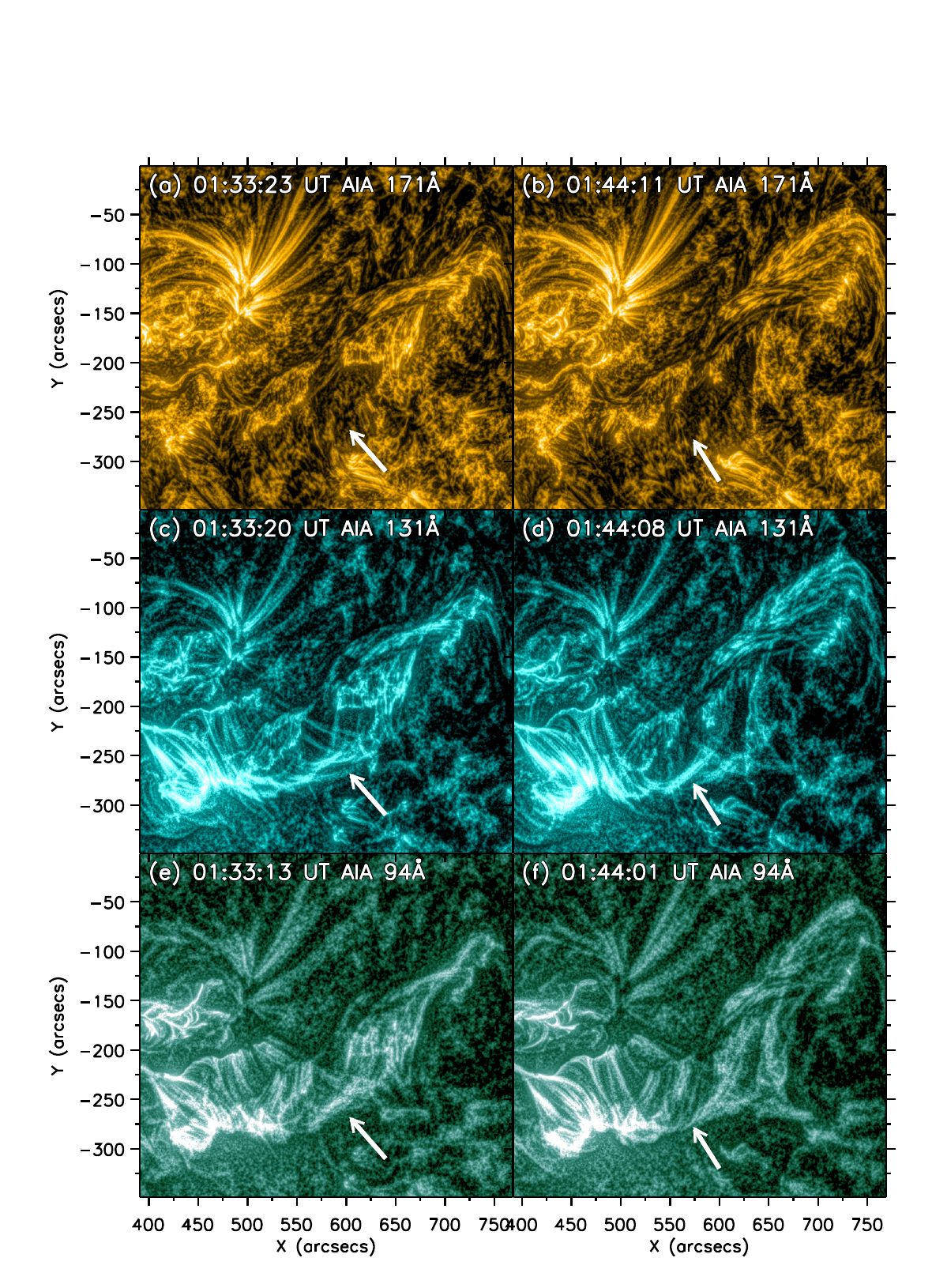}
	\caption{
	Plot showing two times during the triggering of the huge oscillation in AIA 171\AA, 131\AA\ and 94\AA. The images have been processed with the WOW technique described in Sect. \ref{sec:datadescription}. 
The left column shows the region at 1:33 UT. In 171\AA\ (a) the structures already discussed in Fig. \ref{fig:multiplanel171-20150315} are visible. In 131\AA\ (c) however, the post-flare loops around the flare are visible, but also bright tubes in this passband from the flare area to the filament. The white arrows indicate the region where the bright tubes appear in 131\AA\ and 94\AA. These appear to trace a twisted field line reminiscent of a magnetic rope. Similar structures can be seen in 94\AA\ (e). Later, in the right column, the structure has evolved, but the bright tubes connecting to the prominence at both wavelengths are still clear. Note that these bright tubes are not visible in 171\AA. In 171\AA\ the white arrows show the areas where the tubes appear in 131\AA\ and 94\AA.
An animation of this figure is available  \href{}{online}. 
	\label{fig:multiplanel-94-131-171-20150315}}
\end{figure*}
Looking again at Fig.~\ref{fig:multiplanel171-20150315} {and the associated movie}, we cannot identify any plasma emitting in the 171\AA\ passband clearly pushing the filament's cold plasma.  
To determine whether there is hotter plasma pushing in, we study hotter lines such as 131\AA\ and 94\AA. 
Fig. \ref{fig:multiplanel-94-131-171-20150315} shows {snapshots taken in  the 131\AA\ and 94\AA\ passbands at} two times during {the} triggering {phase}. %\sout{at 131\AA\ and 94\AA}
We have also plotted 171\AA\ images {(top panel row)} to give them context as the RF environment is better visualized {in them}.
In these images, we have employed the WOW technique (see Sect. \ref{sec:datadescription}) to enhance the visualization of the coronal structures.
On the left side of the images {in the hotter channels}, we can clearly see the bright postflare loops (Fig. \ref{fig:multiplanel-94-131-171-20150315}(c) to \ref{fig:multiplanel-94-131-171-20150315}(f)). 
In the images, the northern ribbon can be seen as a bright thin line where these loops are rooted. {In fact, this} ribbon can be seen in all three channels shown in the figure. However, the southern ribbon is less clear and in 131\AA\ and 94\AA\ is hidden behind the postflare loops.
In the East of this whole structure, the post flare loops form an arcade connecting the two north and south ribbons, but towards the West {at} 
%\sout{in} 
about $x=500$ arcsecs, the arcade is reoriented and the north foot becomes connected to the magnetic structure of the filament.
This plasma traces a twisted field line from the northern ribbon to the prominence.
These hot structures are visible {in the} 
%\sout{at} 
131\AA\ and 94\AA\ {channels} for more than 10 minutes although different tubes appear to be illuminated as can be {checked by comparing the two}
%\sout{contrasted at both} 
times shown in Fig. \ref{fig:multiplanel-94-131-171-20150315}.
The other AIA wavelengths have been analyzed and these structures could not be identified.

\begin{figure}
    \centering
    \includegraphics[width=\sizefactorhalf\textwidth]{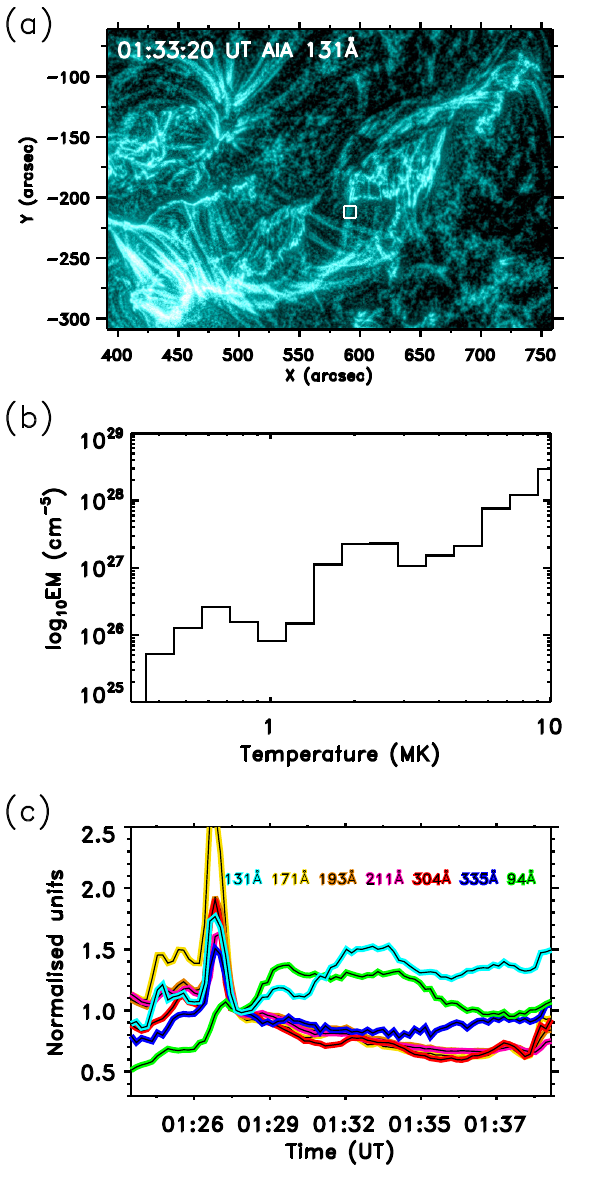}
    \caption{Temperature analysis of the small region shown in the box in (a). In (b) the temperature analysis of that area using DEM is shown. In (c) the average intensity in the same area of all AIA wavebands is shown. The colours used are inspired by the respective AIA channel palette.}
    \label{fig:temperature}
\end{figure}

The analysis of the AIA images indicates that there are tubes connecting the flare zone to the filament that may contain very hot plasma.
Fig. \ref{fig:temperature} shows an analysis of the temperature of one of those flux tubes. 
In panel \ref{fig:temperature}(a), we show a small box where we analyze the temperature distribution at 1:33 UT using the differential emission measure (DEM) inversion technique proposed by \citet{Hannah2012}. This DEM method effectively constrains the emission measure (EM) values using the AIA/SDO observations in six wavelength channels (94, 131, 171, 193, 211, and 335 \AA). Here, we used a temperature range from log T(K) = 5.5  -- 7
with 15 different bins of width $\Delta$log T = 0.1.
It can be seen {(panel b)} that there are two peaks 
%\fminote{say 'broad peaks' instead of just 'peaks'?} 
in the temperature distribution, one around 0.6MK and the other at 2MK. 
However, the EM distribution increases for higher temperatures, {with the maximum EM in the figure being reached at $T=10$~MK, where the technique loses its validity.} 
%\sout{reaching 10MK where the technique is no longer valid.}
%
This distribution reveals the existence of very hot plasma in these tubes. 
We are cautious with this statement as recent studies show that the DEM inversion technique overestimates the presence of hot plasma ($T>2.5MK$) by a factor of 3 to 15 \citep{Athiray_Can-emission-me_2024}.
{The DEM analysis allows for the estimation of electron density and gas pressure \citep[see, e.g.] [among others]{saqri_differential_2020,paraschiv_thermal_2022}. However, in our case,
{the EM may well be overestimated in the hotter part of the range shown in Figure~\ref{fig:temperature}b,}
%\sout{for the 10MK range the EM is largely overestimated} 
which would lead to an erroneous calculation of these parameters. In addition, there are other sources of uncertainty. {The cited papers} assume that the emitting plasma is cylindrical and that the amount of plasma emitted along the line of sight (LOS) coincides with the thickness of the tube seen in the images. In our case, we see a well-defined tube but there is probably much more hot plasma along the LOS. Furthermore, the filling factor, which cannot be measured, contributes to the uncertainty. This makes {the DEM method} unreliable for a quantitative determination of density and pressure.  
Despite all this, the EM indicates the presence of very hot plasma, {at least}10 MK, as suggested by the intensities in the different passbands.}
In panel (c), the temporal sequence of the average intensity for all AIA EUV channels within the small box area considered. From 1:27 to 1:38 UT the intensity in 94 and 131\AA\ increases whereas it decreases for the rest of the AIA channels.
This agrees with the existence of very hot plasma. 
According to the temperature response functions for channels 131\AA\ and 94\AA\ both have two emission peaks. In both cases, the first {one} is below 2MK {whereas the second one} 
%\sout{and the second} 
is above 6.3MK.
If the temperature of the plasma were below 2MK, one would expect emission in channels like 171\AA\ for example. However, clear emission is only observed in 131\AA\ and 94\AA.
These two lines only emit in isolation at around 10MK, which indicates that the plasma in these tubes must have a high temperature.
After 1:38 UT the bright tube that crosses the small box is no longer visible. However, other bright tubes appear as can be seen in Fig. \ref{fig:multiplanel-94-131-171-20150315}(d). This indicates that heating is occurring {in} 
%\sout{at} 
different field lines as the flare evolves.

\begin{figure}[!h]%\vspace{-1cm}\hspace{-0.5cm}
	\vspace{-0.2cm}%\hspace{-0.5cm}
 \centering
	\includegraphics
	[width=0.4\textwidth]
	{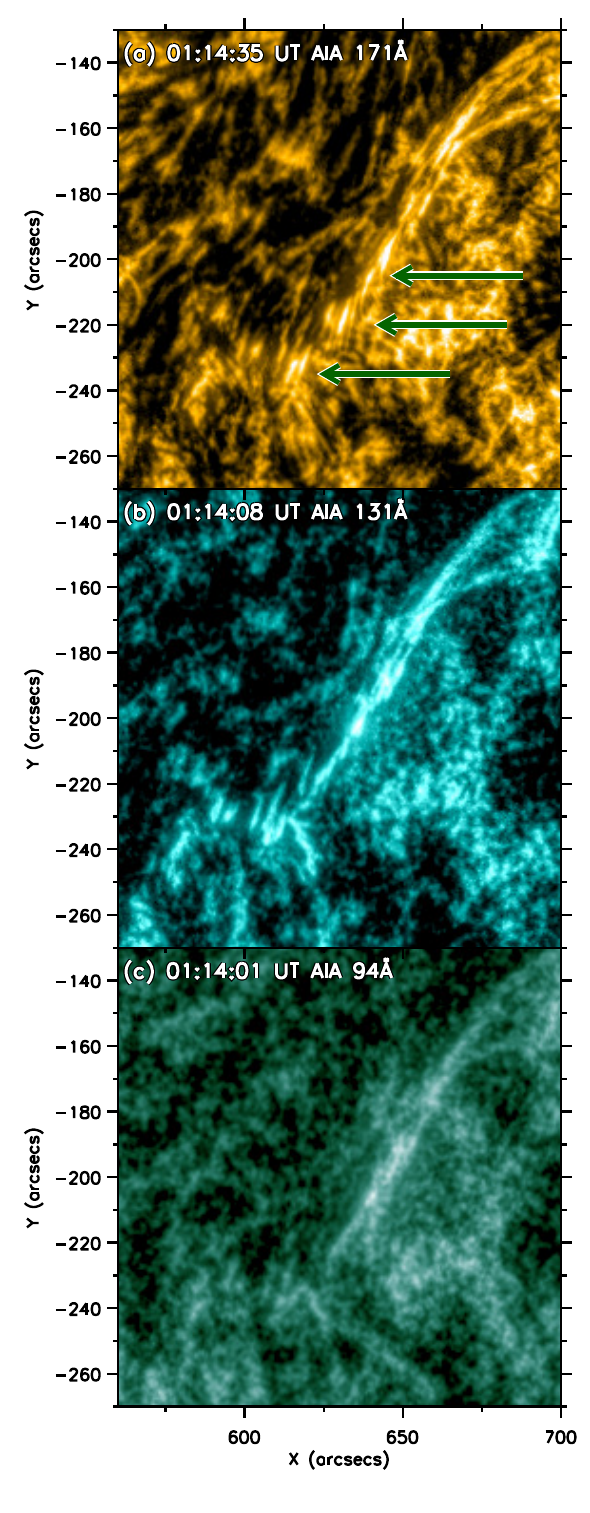}
	\caption{Details of the brightening observed in the prominence in AIA 171\AA, 131\AA\ and 94\AA\ passbands. The region corresponds to the white box in Fig. \ref{fig:multiplanel171-20150315}(d). The green arrows show the bright fibril-like structures at the southern edge of the prominence.\label{fig:fibrils-multiplanel-94-131-171-20150315}}
\end{figure}

During the impulsive phase, the prominence plasma becomes brighter in 171\AA\ (see Fig. \ref{fig:multiplanel171-20150315}) as well as in the 131\AA\ and 94\AA\ bands, as seen in Figs. \ref{fig:multiplanel-94-131-171-20150315}(c) and \ref{fig:multiplanel-94-131-171-20150315}(e). These brightenings may be due to the interaction of the hot plasma of the flare with the cold plasma of the prominence. 
Fig. \ref{fig:fibrils-multiplanel-94-131-171-20150315} shows a detail of the observed fibrilar structures in 
\ifnum 2<1
It should be noted that the figure does not show a temporal evolution. Instead, they show two times on the left and right where the structures are as clear as possible. 
The left column
The figure contains
\fi
a region at the southern end of the prominence (marked by {the} %\sout{a}
white box in Fig. \ref{fig:multiplanel171-20150315}(d)). In 171\AA\ {(Fig.~\ref{fig:fibrils-multiplanel-94-131-171-20150315}a),}  the green arrows point to an area where fibrilar structures are visible. They are also visible{, and with a similar shape,} in 131\AA\ {(panel b)}. %\sout{and are similar in shape to those seen in 171\AA}.  
In 94\AA\ {(panel c)}, a very faint structure is discernible {in those positions}, although it is difficult to associate {it} with fibrils.
\ifnum 2<1
In the right column are the chromospheric fibrils on one of the ribbons (white box in Fig. \ref{fig:multiplanel171-20150315}(e)). \fminote{((chromospheric fibrils in 171, 131 and 94 \AA\ ??))} These are very clear in 171\AA\ and can be seen under the foreground emission in 131\AA, but, they are not so clear in 94\AA. These structures are well-known and have already been studied by REFS\mlnote{(To Fernando, Do you know of any reference to these fibril-like structures in flare ribbons?)}\fminote{(no, I am afraid not)}.
This suggests that filament brightenings may be associated with the same mechanism that also produces chromospheric brightenings.
\fi

Based on this figure and the preceding ones, it is evident that the brightenings predominantly appear in the southern section of the filament, situated in closer proximity to the flare. The threads exhibit increased luminosity towards their ends, where they are closest to the flare and experience the impact of the hot plasma.

\section{Some estimates}\label{sec:theoretical-analysis}

This analysis suggests a possible mechanism for the triggering of the huge oscillation of the RF segment.
The hot plasma is only on one side of the filament, in the SE, and is only visible during the impulsive phase of the oscillation. 
This hot plasma increases the pressure on this side of the filament, pushing the filament to the NW and initiating the oscillation.
This hot plasma is possibly chromospheric plasma evaporated in the ribbons in the flare zone. Due to the energy and mass supply of the flare, we think that the pressure and temperature remain high during the impulsive phase.

We have developed a relatively simple model (see Appendix A) that may represent the temperature increase at one leg due to the energisation process which eventually produces the excitation of the large oscillation. Using this model we have obtained some estimates associated with the pressure increment and the subsequent relaxation. Although a detailed analysis of the process requires the numerical solution of the MHD equations that include the effects of conduction, radiation and heating together with a model of the chromosphere, we have investigated the use of a simplified model that allows us to obtain some analytical expressions for the displacement of the threads and their maximum amplitude of oscillation.

The model consists of a flux tube of length $L$ with a central part of curvature radius $R$. Details of the geometry are given in Appendix A.
We consider a thread of length $\lthread$ and density $\rho_{\rm t}$. The rest of the tube outside the thread has a density $\rho_0$. 
The thread is initially located at $L/2$ which represents the bottom part of the dip. Due to an overpressure at the left leg (from $p_0$ to $p_{\rm max}$) the thread achieves a new equilibrium position at $L/2+\delta s_{\rm e}$. The thread displaces to the right side and upwards as it moves along the curved dip.
According to our linear calculations, this displacement should be of the order of (see Eq.~\eqref{eq:displA})
\begin{align}
    \delta s_{\rm e}=\frac{1}{\gamma}\cfrac{p_{\rm max}/{p_0}-1}{\cfrac{g \, \rho_{\rm t} \, \lthread}{R \rho_0 c_{s0}^2}+\cfrac{2}{L-\lthread}},\label{eq:displ}
\end{align}
where $\gamma=5/3$. The numerator of this equation shows that the displacement of the thread to a new equilibrium location is linearly proportional to the pressure increment.

Once the thread is in its new equilibrium position we assume that the overpressure at the leg suddenly disappears going from $p_{\rm max}$ to $p_{0}$. Therefore, the thread falls back to the bottom of the dip at $L/2$ and oscillates around this position with an amplitude $\delta s_{\rm e}$, because it has gained some extra energy due to the initial rise in pressure. Using energy conservation of the system composed of the dense thread and the evacuated parts, the maximum velocity of the  oscillating thread is found to be (see Eq.~\eqref{eq:veloA})
\begin{equation}\label{eq:velo}
    v=\delta s_{\rm e} \sqrt{\frac{g}{R}}
\end{equation}
where $\delta s_{\rm e}$ is given by Eq.~\eqref{eq:displ}. 
We aim to compare these values for the thread displacement and velocity with the ones derived from the observations to assess the scenario of the oscillation induced by an initial overpressure followed by the relaxation.

From the analysis of the AIA images in 171\AA\ and 131\AA\ we have estimated that $L \approx 620$ arcsecs and that $\lthread$ is between 10 and 20 arcsecs.  Note that those values are the plane of the sky projections.
Furthermore, from the previous section, we saw that the thermal increment in the left foot can exceed 10 MK. Thus we assume that $p_{\rm max}/p_0 \sim 10$, where the initial temperature is considered to be 1 MK.
The radius of curvature is taken as the value obtained from the seismological estimation of Table \ref{table:1}, $R=159\Mm$.
With these numbers and from Eq. \eqref{eq:displ} we estimate a range of values for $\delta s_{\rm e} = 75-38$ Mm where the first value corresponds to $\lthread=10$ and the second to 20 Mm. This range of values comprises the value obtained in the oscillation amplitude $\amplitudelalosecondphase$. 
Similarly, from Eq. \eqref{eq:velo} we obtain a $v=98-51 \kms$ range. The value obtained with the oscillation fit is $\velocitylalosecondphase$, in agreement too with the estimation.
The values of the rough estimates indicate that the proposed mechanism of temperature and pressure increase on one side of the filament can explain the oscillations quantitatively.

This simple model provides also an estimate of the energy released by the flare. The value obtained is a lower limit since only part of the energy released reaches the filament.
The flare pushes the filament upwards along the dipped magnetic field lines, converting some of its energy into kinetic and potential energy of the threads. However, the flare also increases the internal energy of the evacuated regions on either side of the flux tube. Additionally, the pressure within the flux tube rises to a peak value ($p_{\rm max}$), and the region right of the filament gets compressed. By accounting for these contributions, we arrive at Eq. (\ref{eq:flare}).
Using some estimates and typical values for the parameters, such as $p_{\rm max}/p_{0} \approx 10$, $\rho_{\rm th}/\rho_{0} = 100$, $p_{0} = 1 \,{\rm dyne \, cm^{-2}}$, and $\rho_{0}/m_p = 10^{8} \,{\rm cm^{-3}}$ (where $m_{\rm p}$ is the proton mass), we can estimate the flare energy:
\begin{equation}
\label{eq:flare-numbers}
E_{\rm flare} \approx 10^{30} \left( \frac{M_{\rm prom}}{10^{14} \,{\rm g}} \right) \,{\rm erg} \, ,
\end{equation}
Here, we considered a filament length ($l_{\rm th}$) between 10 and 20 arcseconds. However, all these values yield identical orders of magnitude. For a typical prominence mass ($M_{\rm prom}$) of $10^{12} - 10^{15}$ g \citep[as cited in][]{Labrosse2010}, the estimated flare energy falls within the range of $10^{28} - 10^{31}$ erg.
The obtained range is broad, encompassing values similar to X-class flare energies \citep[see reviews by][]{benz_flare_2008,shibata_solar_2011}. This indicates that we are overestimating this energy. Our analysis assumed all the mass moves at the maximum velocity; however, only a fraction likely reaches this speed.
An in-depth study with more realistic models and numerical simulations is needed.

\section{Discussion and Conclusions}\label{sec:conclusions}
In this work, we study the oscillations accompanying the two-step filament eruption that occurred on 14-15 March 2015. This event is known as the St Patrick's event of 2015 because the part of the filament that erupted produced a CME and a large geomagnetic storm on Earth on March 17th. 
The eruption occurs in two phases and {was} %\sout{it is}
well studied by \citet{Chandra2017}. In the first phase on March 14th, a small flare and an associated jet, jet1, produce the detachment of part of the filament. 
Both the detached filament (DF) and the remaining filament (RF) segments oscillate. It is unclear from the observations {what} %\sout{which}
is the triggering {agent} of the DF oscillations. However, during this first phase, the DF rises from its original {location} %\sout{position}
to a new metastable position \citep{Chandra2017}. This change in position is likely to perturb the DF plasma, causing its longitudinal oscillation.
The DF segment moves with a longitudinal oscillation of period $\periodlalofirstphaseDF$  and characteristic damping time $\dampingtimelalofirstphaseDF$. 
We apply seismological techniques to obtain a radius of curvature of the dips of the filament of $\radiuslalofirstphaseDF$ and a  magnetic field strength of $\magneticlalofirstphaseDF$.
In the case of the RF segment, part of the jet1 flows reach the filament and trigger its oscillation. From the time-distance analysis, we can see that the jet1 flows propagate with a velocity of $\speedjetone$ reaching the filament and producing a displacement of the plasma of the prominence. The jet1 and the displacement of the prominence are consecutive, so we conclude that it is the impact of the jet flows that is responsible for the initiation of the oscillation. The interaction resembles the one described in \citet{joshi_interaction_2023} which fits the \citet{Luna2021} model for the jet-prominence interaction. 
In this case, only a small part of the filament oscillates with a period of $\periodlalofirstphaseRF$, a damping time of $\dampingtimelalofirstphaseRF$, and a velocity amplitude of $\velocitylalofirstphaseRF$ which is also a LALO.
Using seismological techniques, we infer a radius of curvature of the dipped field lines of $\radiuslalofirstphaseRF$ and a magnetic field strength of $\magneticlalofirstphaseRF$.

In the second phase on March 15th, %\sout{first,} 
another small flare is produced and an associated jet is launched, jet2%\sout{, and}
{; this is followed by}  a huge oscillation %\sout{occurs} 
in the RF segment. The oscillation is a LALO with a period $\periodlalosecondphase$, a damping time of $\dampingtimelalosecondphase$, and a velocity amplitude of $\velocitylalosecondphase$. With these data, we estimate the radius of curvature to be $\radiuslalosecondphase$ and the field to be $\magneticlalosecondphase$. 
We have investigated the possible triggering of this oscillation. Following a detailed analysis, we have found that the impact of jet2 and the triggering of the huge oscillation are not truly consecutive events. In addition, jet2 impacts a small RF area even though the whole filament is set to oscillate. 
This leads us to discard the impact of the jet2 flows as the triggering of the LAOs in the RF.
Using the SDO/AIA images we can see that after jet2 there are structural changes in the filament probably associated with the DF eruption. Around 01:08UT we start to see brightenings in the filament that will later become {massively present} throughout the entire RF. The time-distance analysis shows that the appearance of these brightenings is simultaneous with the triggering stage. This phase lasts for about an hour {after which} %\sout{and then} 
the RF is %\sout{oscillating} 
{seen to oscillate}. 
%\fminote{would it be ok to give here pointers to figure~\ref{fig:timedistance-march15}? perhaps also to specific times, like 1:10 UT to 2:10 UT or similar?}
%
We have analysed the AIA data at 171\AA, 94\AA\ and 131\AA\ and do not observe plasma flows moving from the AR region to the filament. This indicates that it is not a jet-type triggering where the plasmas emanating from the reconnection are pushing the filament cold plasma.
However, images at 131\AA\ and 94\AA\ show flux tubes connecting the prominence to the flare region (Fig.~\ref{fig:multiplanel-94-131-171-20150315}). This indicates that the same mechanisms that heat the post-flare loops also heat the nearby filament channel feet. The bright tubes trace a complete flux rope with the presence of a dip where the prominence is located.
This causes an increase in the pressure on the SE side of the filament, which pushes the filament towards the NW. This heating and pressure increase is not visible in the AIA channels. This heating leads to evaporation of the chromospheric plasma and fills the tubes with hot, dense plasma. It is at this point that the tubes are visible, which is a time after the first moments of triggering.  For this reason, we do not see any agent pushing the filament, as it is initially not visible in the AIA channels.
Another evidence of this possible heating is that the threads of the filament become bright in 171\AA\ and 131\AA. Initially, fibril-like structures appear, which we think are the ends of the threads that are heated by this hot plasma. These structures are similar to those seen in the ribbon, which must be associated with the heating of the chromospheric material by the flare.

We infer that the triggering of the large oscillation observed on 15 March 2015 might be linked to plasma energization during the flare. However, the specific mechanism responsible for transferring flare energy to the filament remains elusive. 
The two structures, while seemingly independent, exhibit the presence of very hot magnetic tubes connecting them. This suggests a magnetic connection between the filament channel and a region near the flare. 
It is plausible that heated plasma from the flare exerts a force on the filament, initiating the oscillation. Three potential mechanisms can be envisioned: (1) energy transfer from the flare to the filament channel at their footpoints, {(2) inclination of the whole filament structure, or (3)} reconnection of the erupting DF filament with the remaining channel.
In the first scenario, the footpoints of the post-flare arcade should be situated close to those of the filament channel. This expectation arises from the fact that the DF belongs to the same channel. According to the CSHKP model \citep{Carmichael1964,Sturrock1966,Hirayama1974,Kopp1976}, the footpoints of post-flare loops experience heating due to particle and heat flows that trigger chromospheric evaporation.
Recent studies by \citet{druett_chromospheric_2023,druett_exploring_2024}, based on 2.5D flare simulations, demonstrate that evaporation flows can potentially leak out of the post-flare arcade into surrounding regions. 
{In Fig. 6 from \citet{druett_chromospheric_2023} the authors show that the evaporation precedes the arrival of the electron beams, in what they call pre-heated loops. Fig. 14 from \citet{druett_exploring_2024} shows 1D cuts along a field line as functions of time. They show that the evaporation starts approximately 10 seconds before that line is reconnected.}%{\bs Which line?}
The authors suggest that this may be due to compressive heating of neighbouring flux tubes or heat diffusion. Notably, evaporation precedes electron beams. This implies that plasma evaporated as a consequence of the flare might escape the arcade and enter the filament channel through its neighbouring footpoints. 
This influx of high-pressure, high-temperature plasma could then push the prominence, triggering the large-amplitude oscillation.
{However, this evaporation appears to occur only on field lines just ahead of the ribbon front and shortly after reconnection, effectively bringing the evaporation just seconds ahead of the reconnection.}
{In the second scenario, the DF eruption produces the elevation of the south-east zone of the RF filament channel resulting in the inclination of the structure. This causes the cold plasma to slide down along the channel to a lower position, which would initiate oscillation around a new equilibrium position. However, this scenario cannot explain the massive brightenings as seen in Fig. \ref{fig:multiplanel171-20150315} or the bright fingers shown in Fig. \ref{fig:fibrils-multiplanel-94-131-171-20150315}, both associated with a temperature increase.}
\begin{figure*}
    \centering
    \includegraphics[width=1\textwidth]{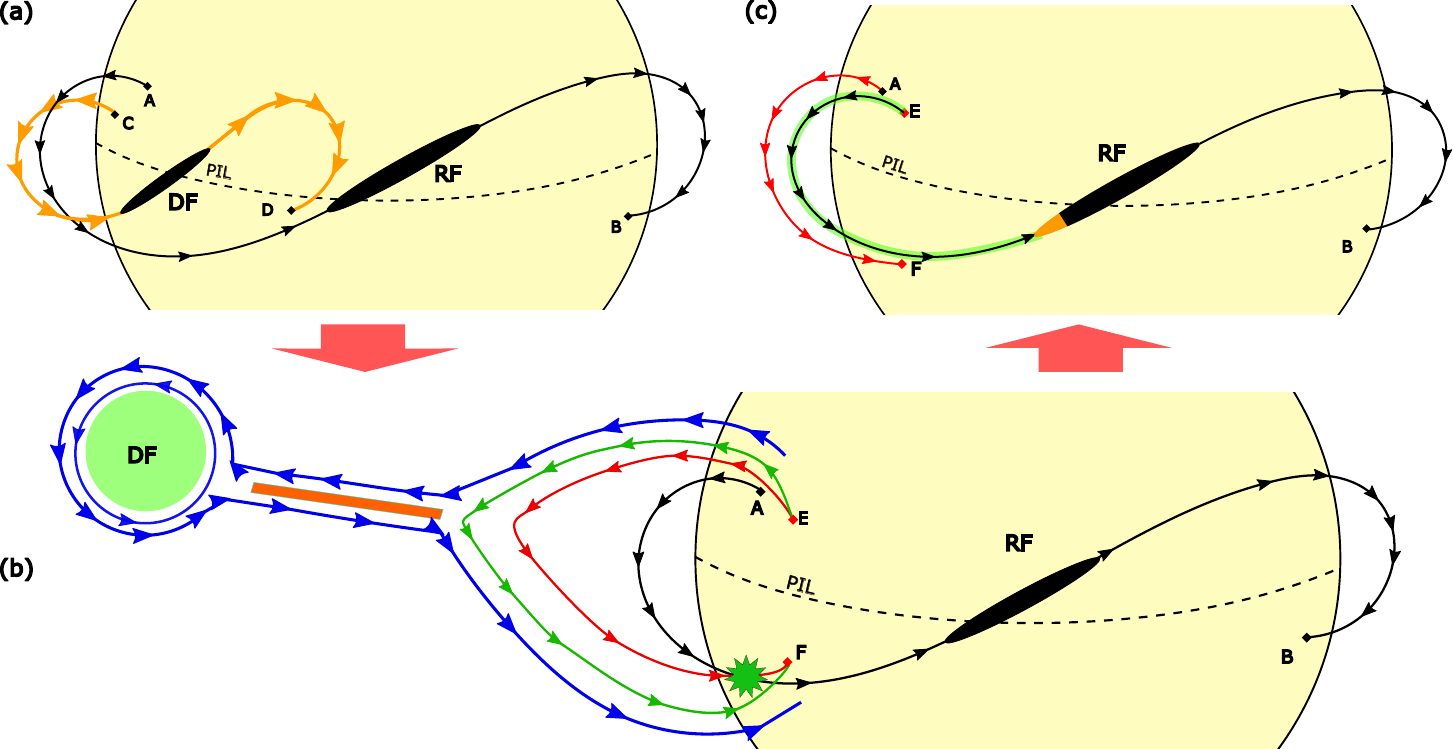}
    \caption{This figure shows a potential mechanism for energy transfer from a solar flare to a filament. In (a) the pre-eruptive state on 14 March after the jet1 is shown. The black solid line represents the field lines for the RF segment whereas the orange line represents the field of the DF segment. Both segments are represented by black ellipsoids. The dashed line is the PIL.
In (b) the DF eruption (green circle) and its magnetic structure (blue lines) are represented. The orange band is the reconnection zone under the erupted filament according to the CSHKP model.
The green and blue lines represent the same field line in different phases. Both connect the points E and F. The green one is newly reconnected in the current layer of the orange band, it contracts and forms the red line. In this contraction, it will reconnect with the line of the filament channel AB. The region where they reconnect is marked by a green star.
In the post-reconnection evolution (c), the change in magnetic field line connectivity is observed. The filament's original footpoint A shifts to E, forming a new footpoint EB and a post-flare loop AF. Hot plasma transferred from the flare is represented by the greenish region east of the EB line, while the orange area on the filament indicates the subsequent brightening.}
    \label{fig:sketch}
\end{figure*}
To illustrate the third scenario, we present a schematic diagram in Fig.~\ref{fig:sketch} on March 15, when the DF and RF segments have already separated.
Note that the structure's size has been intentionally exaggerated relative to the size of the Sun. This allows for a better perspective and by including the limb we give information about the height of the different parts of the structure.
In panel \ref{fig:sketch}(a), two field lines (black lines) depict the filament channel located above the polarity inversion line (PIL) shown as a dashed line on the solar surface.
The line with feet at A and B illustrates the field hosting the RF filament segment, while the line with feet at C and D hosts DF. The RF and DF segments are represented by two black ellipsoids, aligned with the field lines.
Fig. \ref{fig:sketch}(b) shows the eruption of DF. Note that the field line footpoints C and D are no longer shown to simplify the figure.
According to the standard CSHKP model, a reconnection zone, represented by an orange band in panel (b), forms beneath the erupting structure within a current sheet. The core of the coronal mass ejection (CME) is represented by a green circle and contains part of the DF mass. A circular blue line represents the erupting flux rope.
The green curve represents a newly reconnected field line in the current layer, connecting points E and F. This line contracts due to magnetic tension, transforming into the red line. Both lines depict two stages of the same field line evolution and share the same footpoints.
During contraction, the reconnected line (red) intersects with the filament channel lines (black), triggering another reconnection event. This reconnection region is marked by a green star in the figure.
Magnetic reconnection modifies the initial connectivity between the contracting loop EF and the channel line AB. 
The mutual orientation of the reconnecting field lines is unlikely to be antiparallel, given their common origin associated with DF and RF, respectively, but even a comparatively small relative angle may lead to component reconnection with a guide field, which could suffice to explain the change of connectivity and the possibility of exchange of mass and energy along the newly formed, reconnected, field line.
Fig. \ref{fig:sketch}(c) depicts the post-reconnection configuration. Reconnection rearranges the field lines such that point A is now connected to F, while E connects to B. Consequently, the filament channel anchored at footpoint B becomes magnetically linked to the flare region (F).
Furthermore, the hot plasma previously confined within the loop EF now resides to the east of the filament. This hot plasma is illustrated within the tube as a greenish region enveloping the eastern segment of the reconnected field line originating at E. The pressure exerted by the hot plasma on the cooler prominence plasma drives it westward towards footpoint B. The south-west end of the filament, highlighted in orange, represents the brightening observed during the impulsive phase of the oscillation, likely due to the interaction between the hot flare plasma flows and the prominence material. The panel (c) sketch illustrates the structures that appear in the 131\AA\ and 94\AA\ images of Fig. \ref{fig:multiplanel-94-131-171-20150315}.

In this paper, we aim to shed light on the mechanisms behind the triggering of large amplitude oscillations. We find that the very energetic event of 15 March is associated with a nearby flare. The mechanism is neither related to Moreton waves nor to jets as usually reported in the literature. Actually, the energy of the flare is somehow transferred to the filament channel. This increases the pressure on one side of the filament channel, pushing the heavy prominence plasma and initiating the oscillation.
Further cases need to be reported to provide insight into these mechanisms. Realistic numerical experiments are also required to understand the scenarios that produce these events. These investigations will be left for future work.

\begin{acknowledgements}
This publication is part of the R+D+i project PID2020-112791GB-I00, financed by
MCIN/AEI/10.13039/501100011033.
M.\ Luna acknowledges support through the Ram\'on y Cajal fellowship RYC2018-026129-I from the Spanish Ministry of Science and Innovation, the Spanish National Research Agency (Agencia Estatal de Investigaci\'on), the European Social Fund through Operational Program FSE 2014 of Employment, Education and Training and the Universitat de les Illes Balears.
This research has been supported by the European Research Council through the Synergy grant No. 810218 (“The Whole Sun,” ERC-2018-SyG) and also by the Research Council of Norway through its Centres of Excellence scheme, project number 262622. 
CHIANTI is a collaborative project involving George Mason University, the University of Michigan (USA), University of Cambridge (UK) and NASA Goddard Space Flight Center (USA). We acknowledge NASA's Astrophysics Data System Bibliographic Services.

\end{acknowledgements}

\begin{appendix}
\section{Considerations on thread dynamics and energetics}\label{sectapp}
We use a modified version of the model described in \citet{adroveretal2021} to show the basic physics behind the energization of the thread and the subsequent relaxation phase that involves the large amplitude thread oscillations.

We consider the prominences to be composed of magnetic flux tubes with dense cold plasma threads in them. On both sides of the cold thread, the tube has a coronal temperature and is often referred to as the evacuated regions. The flux tubes have dips in their central part and an equilibrium situation the threads will reside at the bottom of these dips. The movement of the threads is along the tube and can be modelled as a 1D system. The main equation of our system is the momentum equation
\begin{eqnarray}\label{momentum}
\rho_{\rm t} \frac{d^2 s}{dt^2}= -\frac{dp}{ds}(s)+\rho_{\rm t}\, g_\parallel(s),
\end{eqnarray}
being $s$ the coordinate of the centre of the thread along the fluxtube, $\rho_{\rm t}$ the thread density and $g_\parallel$ the gravity projection along the magnetic field line.

We approximate the gas pressure derivative by
\begin{eqnarray}\label{dpdsaprox}
\frac{dp}{ds}(s) \approx \frac{\Delta p}{\Delta s}(s)=\frac{p_2(s)-p_1(s)}{\lthread},
\end{eqnarray}
where we have assumed  that the
thread does not change its shape and keeps its length, $\lthread$, constant. Let us suppose that on the left part of the tube, denoted by the subindex 1, and due to the energetisation process gas pressure is increasing and reaches a maximum value $p_{\rm max}$, i.e. $p_1=p_{\rm max}$. This pressure excess pushes the thread to the right and produces a compression on the evecuated part, denoted by the subindex 2. We assume that the compression of this part of the tube, which initially had a pressure $p_0$, is adiabatic, therefore we have that  
\begin{eqnarray}\label{adiabatic1}
p_2(s)=p_0 \left(\frac{L/2-\lthread/2}{L-s-\lthread/2}\right)^\gamma.
\end{eqnarray}
Note that if $s=L/2$ the thread is its initial equilibrium position, and $p_2=p_0$. We aim at finding the new position equilibrium due to the supplied pressure excess. 
Writing $s=L/2+\delta s_{\rm e}$ the previous expression is approximated by 
\begin{eqnarray}\label{adiabatic1b}
p_2(s)\approx p_0 \left(1+2 \gamma \frac{\delta s_{\rm e}}{L-\lthread}\right),
\end{eqnarray}
valid for the situation $\delta s_{\rm e} \ll {L-\lthread}$. Using this linear approximation we have that
\begin{eqnarray}\label{dpdsaproxnew}
-\frac{dp}{ds}\approx \frac{c_{s0}^2}{\gamma}\frac{\rho_0}{\lthread}\left(p_{\rm max}/p_0-1-2 \gamma \frac{\delta s_{\rm e}}{L-\lthread}\right),
\end{eqnarray}
where we have used the definition of the coronal sound speed, $c_{s0}^2=\gamma p_0/\rho_0$.

For the gravitational term in Eq. \eqref{momentum} we consider the approximated expression from \citet{luna_extension_2022} where it is assumed that the dips have a circular geometry of radius $R$
\begin{eqnarray}\label{eq:grav}
g_\parallel\approx -\frac{g}{R_\mathrm{eq}}  \delta s_{\rm e}=-g \left(\frac{1}{R} +\frac{1}{R_{\odot}} \right) \delta s_{\rm e} \, ,
\end{eqnarray}
and $R_{\odot}$ is the solar radius.
From the equilibrium condition we have 
\begin{eqnarray}\label{eq:equil}
0= -\frac{dp}{ds}+\rho_{\rm t}\, g_\parallel,
\end{eqnarray}
and using the previously derived linear approximations we obtain the following thread displacement
\begin{align}
    \delta s_{\rm e}=\frac{1}{\gamma}\cfrac{p_{\rm max}/{p_0}-1}{\cfrac{g}{R_{\rm eq}}\cfrac{\rho_{\rm t}}{\rho_0}\cfrac{\lthread}{c_{s0}^2}+\cfrac{2}{L-\lthread}} \, .\label{eq:displA}
\end{align}
When moving along the dip field line, the thread increases its height with respect to the bottom of the dip which is approximately
\begin{align}\label{eq:deltah}
\Delta h=\frac{1}{R}\frac{\delta s_{\rm e}^2}{2} \, .
\end{align}
Here it should be noted that \eqref{eq:equil} condition gives a lower value for the $\delta s_{\rm e}$ \eqref{eq:displA} displacement. This is because the thread would overshoot the point where the forces cancel out. The condition to find $\delta s_{\rm e}$ would be the point where the velocity becomes zero. However, it is not possible to find an analytical solution and we consider Eq. \eqref{eq:displA} a reasonable approximation or a lower value.
Just at the position of maximum displacement is when the system has the maximum energy. The pressure exerted by the flare is still $p_{\rm max}$ and the total energy of the system is the sum of the internal energies of the evacuated parts of the tube  plus the potential energy of the thread at height $h$ (since the internal energy of the dense thread does not change it is not taken into account in the computation), namely,
\begin{eqnarray}\label{eq:total-energy-max}
    E_\text{T, max}&=&\frac{p_{\rm max}}{\gamma-1}\left(L/2 +\delta s_{\rm e}-\lthread/2\right)+\\ \nonumber
         && \frac{p_0}{\gamma-1}\frac{\left(L/2-\lthread/2\right)^\gamma}{\left(L/2-\delta s_{\rm e}-\lthread/2\right)^{\gamma-1}} +\rho_{\rm t} \, g \, \Delta h \,  \lthread \, .
\end{eqnarray}
Each term contains the length of the segment that represents and we have taken into account that the right leg has gone through an adiabatic compression. 

In the second phase of the process proposed in the present paper as the trigger of the oscillations we assume that the overpressure that has displaced the thread a distance $\delta s_{\rm e}$, suddenly disappears and the thread tends to return to its initial position at $L/2$. 
We are interested in the calculation of the maximum velocity while it is oscillating around $L/2$ since this is a parameter that can be compared with the observational data. Energy conservation allows us the determine the maximum velocity. 
In this phase, the pressure drops because we assume that the flare stops injecting hot plasma into the left side of the tube. We assume that the pressure transitions from $p_{\rm{max}}$ to an ambient atmospheric value $p_0$.
Just after the sudden pressure decrease to $p_0$, the total energy of the system is, similarly to Eq. \eqref{eq:total-energy-max},
\begin{align}\label{eq:total-energy}
    E_T=\frac{p_0}{\gamma-1}\left(\left(s-\lthread/2\right)+\frac{\left(L/2-\lthread/2\right)^\gamma}{\left(L-s-\lthread/2\right)^{\gamma-1}}\right)+\rho_{\rm t} \, g \, \Delta h \,  \lthread \, .
\end{align}
This expression is now written in terms of $\delta s_{\rm e}$ using the same approximations as before and reducing to
\begin{align}\label{eq:energy2}
    E_T=\frac{p_0}{\gamma-1}(L-\lthread) \left(1+\gamma \frac{\delta s_{\rm e}}{L-\lthread}\right) +\rho_{\rm t} \, g \, \Delta h \, \lthread.
\end{align}
As the thread returns to its equilibrium position $s=L/2$ located at the bottom of the dip, it acquires kinetic energy associated with its velocity $v$. The left part of the tube is compressed adiabatically. In this case, the total energy is
\begin{align}
E_T^{'}=\frac{p_0}{\gamma-1}(L-\lthread) \left(1+\gamma \frac{\delta s_{\rm e}}{L-\lthread}\right) +\frac{1}{2}\rho_{\rm t} v^2 \lthread \, .
\end{align}
This expression must be equal to Eq.~(\ref{eq:energy2}) because of energy conservation ($E_T=E_T^{'}$), this allows us to obtain the simple expression (valid when $\delta s_{\rm e} \ll {L-\lthread}$)
\begin{align}
    v=\sqrt{2 \,  g \, \Delta h}=\delta s_{\rm e}\sqrt{\cfrac{g}{R}},\label{eq:veloA}
\end{align}
where $\Delta h$ is given by Eq.~(\ref{eq:deltah}).

An estimation of the energy of a solar flare can be calculated using the total energy (\ref{eq:total-energy-max}) and subtracting the energy at equilibrium ($s=L/2$):
\begin{eqnarray}\label{eq:delta-energy}
\Delta E_T&=&\frac{p_{\rm max}}{\gamma-1}\left(L/2 +{\delta s_{\rm e}}-\lthread/2\right)+ \\ \nonumber
&& \frac{p_0}{\gamma-1}\frac{\left(L/2-\lthread/2\right)^\gamma}{\left(L-{\delta s_{\rm e}}-\lthread/2\right)^{\gamma-1}} +\rho_{\rm t} \, g \, \Delta h \, \lthread - \frac{p_0}{\gamma-1}\left(L-\lthread\right) .
\end{eqnarray}
This equation represents the energy contributed by the flare per unit area to each magnetic flux tube. The total number of tubes or filaments is estimated by assuming a typical prominence mass ($M_{\rm prom}$) and dividing it by the mass of an individual filament ($\rho_{\rm th} l_{\rm th}$). With this, we can assess a lower bound of the flare energy as
\begin{equation}\label{eq:flare}
E_{\rm flare} \approx \Delta E_T \, \frac{M_{\rm prom}}{\rho_{\rm th} \, l_{\rm th}} \, .
\end{equation}

\section{Pendulum model for longitudinal solar prominence oscillations}\label{sectapp2}
In this appendix, we briefly describe the pendulum model by \citet{luna_extension_2022}. The longitudinal oscillations in prominences are characterized by the 1D displacements of the constituent threads along the magnetic flux tubes. We assume a low-$\beta$, adiabatic plasma confined within flux tubes of uniform cross-section, aligned along a static magnetic field, with no heating or radiative terms included.
\citet{luna_effects_2012} demonstrated that, in contrast to the usual MHD slow modes, both the gas-pressure gradient and gravity are the restoring forces for these modes. From the ideal MHD equations
\begin{equation}\label{eq:wave-eq}
\frac{\partial^{2} v}{\partial t^{2}} - c_{s}^{2} \frac{\partial^{2} v}{\partial s^{2}} =\gamma g_{\parallel} \frac{\partial v}{\partial s} + v \frac{\partial g_{\parallel}}{\partial s} ~,
\end{equation}
where $c_{s}^{2}=\gamma \, p/\rho$ is the sound speed and $p$ and $\rho$ are the equilibrium gas pressure and density, respectively. The gravity projected along the magnetic flux tube is $g_{\parallel}(s)$.
Assuming a harmonic time dependence, $e^{i\omega t}$, the governing equation becomes
\begin{equation}\label{eq:wave-eq-harmonic}
c_{s}^{2} \frac{\partial^{2} v}{\partial s^{2}} + \gamma g_{\parallel} \frac{\partial v}{\partial s} +\left(\omega^2 + \frac{\partial g_{\parallel}}{\partial s}\right) v =0.
\end{equation}
The novelty introduced by \citet{luna_automatic_2022} is that the intrinsic spatial variations of the solar gravity field are considered $\vec{g}=\vec{g}(\vec{r})$ in contrast with the usual assumption of uniform gravity in Cartesian geometry. 
Additionally, it is assumed that prominences are located at the lower corona where $r \approx \rsun$ being $r$ the radial distance to the solar centre and $\rsun$ the solar radius. Thus, $\vec{g} \approx -g_{0} \, \vec{\hat{r}}$, where $\vec{\hat{r}}$ is the radial unit vector pointing in the (minus) direction of gravity and $g_0 = 274 \kmss$ the gravitational acceleration at the solar surface. This assumes that, for filaments in the low corona, the main spatial variation of gravity is due to its changing direction, not its intensity because the radial distance $r$ is not changing considerably. 
Assuming that dips have a semicircular geometry with radius of curvature $R$ and under some approximations \citep[see Sect. 3.1 from][]{luna_automatic_2022} the projected gravity is
\begin{equation}\label{eq:approximated-gravity}
 g_\parallel = -\grav \, \left(\frac{1}{R} + \frac{1}{R_\odot}\right) \, s  \,.
\end{equation}
Note that on the right-hand side, the first term is associated with the change of magnetic-field direction along the line and the second is associated with the intrinsic variation of nonuniform gravity.
Substituting Eq. \eqref{eq:approximated-gravity} into Eq. \eqref{eq:wave-eq-harmonic} we obtain a dispersion relation that can be approximated as
\begin{equation}\label{eq:frequency-approximation}
\omega^{2} \approx \frac{\grav}{R} + \frac{\grav}{\rsun} + \omega^{2}_\mathrm{slow} \, ,
\end{equation}
where $\omega_\mathrm{slow}$ is the slow-mode angular frequency associated with the gas-pressure gradient. \citet{luna_effects_2012} showed that the last term associated with slow modes is relatively small and can be neglected. Additional works supported this finding \citep[see][]{zhang_parametric_2013,luna_robustness_2016,zhou_three-dimensional_2018,zhang_damping_2019,liakh_large-amplitude_2021}. Neglecting the slow mode contribution, Eq. \eqref{eq:frequency-approximation} can written in terms of the period ($\omega=2\pi/P$) as 
\begin{equation}\label{eq:period-approximation}
\frac{1}{P^{2}}=\frac{\grav}{4\pi^{2}\,R}+\frac{1}{\psun^{2}} \, ,
\end{equation}
where
\begin{equation}\label{eq:period-sun}
\psun=2\pi \, \sqrt{\frac{\rsun}{\grav}}=167 \mins \, .
\end{equation}
This corresponds to the oscillation period of a thread in a straight horizontal tube. In contrast to the uniform gravity case, gravity still has a non-zero projection along the tube.
From Eq. \eqref{eq:period-approximation} we find
\begin{equation}\label{eq:new-pendulum-appendix}
    \frac{R}{R_\odot}=\frac{\left(P/P_\odot\right)^2}{1-\left(P/P_\odot\right)^2} \, ,
\end{equation}
which is a useful equation for determining the radius of curvature of the dip when the oscillation period is known.

This model also predicts a minimum value for the magnetic field strength of the prominence, assuming that the magnetic field tension supports the cool, dense prominence threads. Consequently, the magnetic tension must be greater than the weight of the threads, such that
\begin{equation}\label{eq:magneticsupport-condition}
\frac{B^{2}}{\mu \, R} -\rho \, g_0 \ge 0 \, .
\end{equation}
Eq. \eqref{eq:new-pendulum-appendix} yields the following relation:
\begin{equation}\label{eq:magnetic-field-seismology-appendix}
    B \ge \sqrt{\mu \, \rho\, g_0\, \rsun} \, \frac{P/\psun}{\sqrt{1-\left( P/\psun \right)^2}} \, .
\end{equation}
Using this expression, the magnetic field strength at the dip of the field lines supporting the prominence can be estimated.

% We note that $g_{\parallel}$ depends on the position $s$ due to two contributions: the change in the projection of the gravity along the field line and the intrinsic variation of the gravity field

\end{appendix}

%\bibliography{bibliography}

\begin{thebibliography}{79}
\expandafter\ifx\csname natexlab\endcsname\relax\def\natexlab#1{#1}\fi

\bibitem[{{Adrover-Gonz{\'a}lez} {et~al.}(2021){Adrover-Gonz{\'a}lez},
  {Terradas}, {Oliver}, \& {Carbonell}}]{adroveretal2021}
{Adrover-Gonz{\'a}lez}, A., {Terradas}, J., {Oliver}, R., \& {Carbonell}, M.
  2021, \aap, 649, A142

\bibitem[{{Arregui} {et~al.}(2018){Arregui}, {Oliver}, \&
  {Ballester}}]{Arregue2018}
{Arregui}, I., {Oliver}, R., \& {Ballester}, J.~L. 2018, Living Reviews in
  Solar Physics, 15, 3

\bibitem[{{Asai} {et~al.}(2012){Asai}, {Ishii}, {Isobe}, {Kitai}, {Ichimoto},
  {UeNo}, {Nagata}, {Morita}, {Nishida}, {Shiota}, {Oi}, {Akioka}, \&
  {Shibata}}]{asai2012}
{Asai}, A., {Ishii}, T.~T., {Isobe}, H., {et~al.} 2012, \apjl, 745, L18

\bibitem[{Athiray \& Winebarger(2024)}]{Athiray_Can-emission-me_2024}
Athiray, P.~S. \& Winebarger, A.~R. 2024, {Can emission measure distributions
  derived from extreme-ultraviolet images accurately constrain high temperature
  plasma?}, arXiv:2401.12372 [astro-ph]

\bibitem[{Auch{\`e}re {et~al.}(2023)Auch{\`e}re, Soubri{\'e}, Pelouze, \&
  Buchlin}]{auchere_image_2023}
Auch{\`e}re, F., Soubri{\'e}, E., Pelouze, G., \& Buchlin, {\'E}. 2023,
  Astronomy \& Astrophysics, 670, A66

\bibitem[{{Bamba} {et~al.}(2019){Bamba}, {Inoue}, \& {Hayashi}}]{Bamba2019}
{Bamba}, Y., {Inoue}, S., \& {Hayashi}, K. 2019, \apj, 874, 73

\bibitem[{Becker(1958)}]{becker1958}
Becker, U. 1958, Zeitschrift f{\"u}r Astrophysik, 44, 243

\bibitem[{Benz(2008)}]{benz_flare_2008}
Benz, A.~O. 2008, Living Reviews in Solar Physics, 5, 1, publisher: Springer
  International Publishing

\bibitem[{{Bocchialini} {et~al.}(2011){Bocchialini}, {Baudin}, {Koutchmy},
  {Pouget}, \& {Solomon}}]{bocchialini2011}
{Bocchialini}, K., {Baudin}, F., {Koutchmy}, S., {Pouget}, G., \& {Solomon}, J.
  2011, \aap, 533, A96

\bibitem[{Bruzek \& Becker(1957)}]{bruzek1957}
Bruzek, A. \& Becker, U. 1957, Zeitschrift f{\"u}r Astrophysik, 42, 76

\bibitem[{{Carmichael}(1964)}]{Carmichael1964}
{Carmichael}, H. 1964, in {The Physics of Solar Flares}, ed. W.~N. {Hess}, 451

\bibitem[{{Chandra} {et~al.}(2017){Chandra}, {Filippov}, {Joshi}, \&
  {Schmieder}}]{Chandra2017}
{Chandra}, R., {Filippov}, B., {Joshi}, R., \& {Schmieder}, B. 2017, \solphys,
  292, 81

\bibitem[{{Chen} {et~al.}(2008){Chen}, {Jiang}, \& {Ma}}]{chen2008}
{Chen}, H.~D., {Jiang}, Y.~C., \& {Ma}, S.~L. 2008, \aap, 478, 907

\bibitem[{{Devi} {et~al.}(2022){Devi}, {Chandra}, {Joshi}, {Chen}, {Schmieder},
  {Uddin}, \& {Moon}}]{Devi2022}
{Devi}, P., {Chandra}, R., {Joshi}, R., {et~al.} 2022, arXiv e-prints,
  arXiv:2202.13147

\bibitem[{Dodson(1949)}]{dodson1949}
Dodson, H.~W. 1949, Astrophysical Journal, 110, 382

\bibitem[{Druett {et~al.}(2024)Druett, Ruan, \&
  Keppens}]{druett_exploring_2024}
Druett, M., Ruan, W., \& Keppens, R. 2024, Astronomy \& Astrophysics, 684, A171

\bibitem[{Druett {et~al.}(2023)Druett, Ruan, \&
  Keppens}]{druett_chromospheric_2023}
Druett, M.~K., Ruan, W., \& Keppens, R. 2023, Solar Physics, 298, 134

\bibitem[{Eto {et~al.}(2002)Eto, Isobe, Narukage, Asai, Morimoto, Thompson,
  Yashiro, Wang, Kitai, Kurokawa, \& Shibata}]{eto2002}
Eto, S., Isobe, H., Narukage, N., {et~al.} 2002, Publications of the
  Astronomical Society of Japan, 54, 481

\bibitem[{Foullon {et~al.}(2009)Foullon, Verwichte, \&
  Nakariakov}]{foullon2009}
Foullon, C., Verwichte, E., \& Nakariakov, V.~M. 2009, The Astrophysical
  Journal, 700, 1658

\bibitem[{{Freeland} \& {Handy}(1998)}]{Freeland1998}
{Freeland}, S.~L. \& {Handy}, B.~N. 1998, \solphys, 182, 497

\bibitem[{Gibson {et~al.}(2010)Gibson, Kucera, Rastawicki, Dove, de~Toma, Hao,
  Hill, Hudson, Marqu{\'e}, McIntosh, Rachmeler, Reeves, Schmieder, Schmit,
  Seaton, Sterling, Tripathi, Williams, \&
  Zhang}]{gibson_three-dimensional_2010}
Gibson, S.~E., Kucera, T.~A., Rastawicki, D., {et~al.} 2010, The Astrophysical
  Journal, 724, 1133, publisher: IOP Publishing

\bibitem[{Gilbert {et~al.}(2008)Gilbert, Daou, Young, Tripathi, \&
  Alexander}]{gilbert2008}
Gilbert, H.~R., Daou, A.~G., Young, D., Tripathi, D., \& Alexander, D. 2008,
  \apj, 685, 629

\bibitem[{{Hannah} \& {Kontar}(2012)}]{Hannah2012}
{Hannah}, I.~G. \& {Kontar}, E.~P. 2012, \aap, 539, A146

\bibitem[{Hershaw {et~al.}(2011)Hershaw, Foullon, Nakariakov, \&
  Verwichte}]{hershaw2011}
Hershaw, J., Foullon, C., Nakariakov, V.~M., \& Verwichte, E. 2011, Astron.
  Astrophys., 531, A53

\bibitem[{{Hirayama}(1974)}]{Hirayama1974}
{Hirayama}, T. 1974, \solphys, 34, 323

\bibitem[{Hudson {et~al.}(1999)Hudson, Acton, Harvey, \&
  McKenzie}]{hudson_stable_1999}
Hudson, H.~S., Acton, L.~W., Harvey, K.~L., \& McKenzie, D.~E. 1999, The
  Astrophysical Journal, 513, L83, publisher: IOP Publishing

\bibitem[{{Hyder}(1966)}]{Hyder1966}
{Hyder}, C.~L. 1966, \zap, 63, 78

\bibitem[{Isobe \& Tripathi(2006)}]{isobe2006}
Isobe, H. \& Tripathi, D. 2006, Astron. Astrophys., 449, L17

\bibitem[{{Isobe} {et~al.}(2007){Isobe}, {Tripathi}, \&
  {Archontis}}]{isobe2007}
{Isobe}, H., {Tripathi}, D., \& {Archontis}, V. 2007, \apj, 657, L53

\bibitem[{{Jing} {et~al.}(2006){Jing}, {Lee}, {Spirock}, \& {Wang}}]{jing2006}
{Jing}, J., {Lee}, J., {Spirock}, T.~J., \& {Wang}, H. 2006, \solphys, 236, 97

\bibitem[{{Jing} {et~al.}(2003){Jing}, {Lee}, {Spirock}, {Xu}, {Wang}, \&
  {Choe}}]{jing2003}
{Jing}, J., {Lee}, J., {Spirock}, T.~J., {et~al.} 2003, \apjl, 584, L103

\bibitem[{Joshi {et~al.}(2023)Joshi, Luna, Schmieder, Moreno-Insertis, \&
  Chandra}]{joshi_interaction_2023}
Joshi, R., Luna, M., Schmieder, B., Moreno-Insertis, F., \& Chandra, R. 2023,
  Astronomy \& Astrophysics, 672, A15, publisher: EDP Sciences

\bibitem[{{Kopp} \& {Pneuman}(1976)}]{Kopp1976}
{Kopp}, R.~A. \& {Pneuman}, G.~W. 1976, \solphys, 50, 85

\bibitem[{{Labrosse} {et~al.}(2010){Labrosse}, {Heinzel}, {Vial}, {Kucera},
  {Parenti}, {Gun{\'a}r}, {Schmieder}, \& {Kilper}}]{Labrosse2010}
{Labrosse}, N., {Heinzel}, P., {Vial}, J.~C., {et~al.} 2010, \ssr, 151, 243

\bibitem[{{Lemen} {et~al.}(2012){Lemen}, {Title}, {Akin}, {Boerner}, {Chou},
  {Drake}, {Duncan}, {Edwards}, {Friedlaender}, {Heyman}, {Hurlburt}, {Katz},
  {Kushner}, {Levay}, {Lindgren}, {Mathur}, {McFeaters}, {Mitchell}, {Rehse},
  {Schrijver}, {Springer}, {Stern}, {Tarbell}, {Wuelser}, {Wolfson}, {Yanari},
  {Bookbinder}, {Cheimets}, {Caldwell}, {Deluca}, {Gates}, {Golub}, {Park},
  {Podgorski}, {Bush}, {Scherrer}, {Gummin}, {Smith}, {Auker}, {Jerram},
  {Pool}, {Soufli}, {Windt}, {Beardsley}, {Clapp}, {Lang}, \&
  {Waltham}}]{Lemen2012}
{Lemen}, J.~R., {Title}, A.~M., {Akin}, D.~J., {et~al.} 2012, \solphys, 275, 17

\bibitem[{{Li} \& {Zhang}(2012)}]{Li2012}
{Li}, T. \& {Zhang}, J. 2012, \apjl, 760, L10

\bibitem[{Liakh {et~al.}(2020)Liakh, Luna, \& Khomenko}]{liakh_numerical_2020}
Liakh, V., Luna, M., \& Khomenko, E. 2020, Astronomy \& Astrophysics, 637, A75,
  publisher: EDP Sciences

\bibitem[{Liakh {et~al.}(2021)Liakh, Luna, \&
  Khomenko}]{liakh_large-amplitude_2021}
Liakh, V., Luna, M., \& Khomenko, E. 2021, Astronomy \& Astrophysics, 654,
  A145, publisher: EDP Sciences

\bibitem[{Liakh {et~al.}(2023)Liakh, Luna, \& Khomenko}]{liakh_numerical_2023}
Liakh, V., Luna, M., \& Khomenko, E. 2023, Astronomy \& Astrophysics, 673,
  A154, publisher: EDP Sciences

\bibitem[{{Liu} {et~al.}(2012){Liu}, {Ofman}, {Nitta}, {Aschwanden},
  {Schrijver}, {Title}, \& {Tarbell}}]{liu2012}
{Liu}, W., {Ofman}, L., {Nitta}, N.~V., {et~al.} 2012, \apj, 753, 52

\bibitem[{{Liu} {et~al.}(2015){Liu}, {Hu}, {Wang}, {Yang}, {Zhu}, {Liu},
  {Luhmann}, \& {Richardson}}]{Liu2015}
{Liu}, Y.~D., {Hu}, H., {Wang}, R., {et~al.} 2015, \apjl, 809, L34

\bibitem[{Luna {et~al.}(2012)Luna, D\'{\i}az, \& Karpen}]{luna_effects_2012}
Luna, M., D\'{\i}az, A.~J., \& Karpen, J. 2012, The Astrophysical Journal, 757,
  98, publisher: IOP Publishing

\bibitem[{Luna \& Karpen(2012)}]{luna_large-amplitude_2012}
Luna, M. \& Karpen, J. 2012, The Astrophysical Journal, 750, L1, publisher: IOP
  Publishing

\bibitem[{Luna {et~al.}(2018)Luna, Karpen, Ballester, Muglach, Terradas,
  Kucera, \& Gilbert}]{luna_gong_2018}
Luna, M., Karpen, J., Ballester, J.~L., {et~al.} 2018, The Astrophysical
  Journal Supplement Series, 236, 35, publisher: IOP Publishing

\bibitem[{Luna {et~al.}(2014)Luna, Knizhnik, Muglach, Karpen, Gilbert, Kucera,
  \& Uritsky}]{luna_observations_2014}
Luna, M., Knizhnik, K., Muglach, K., {et~al.} 2014, The Astrophysical Journal,
  785, 79, publisher: IOP Publishing

\bibitem[{Luna {et~al.}(2022)Luna, Mestre, \&
  Auch{\`e}re}]{luna_automatic_2022}
Luna, M., Mestre, J. R.~M., \& Auch{\`e}re, F. 2022, Astronomy \& Astrophysics,
  666, A195, publisher: EDP Sciences

\bibitem[{{Luna} \& {Moreno-Insertis}(2021)}]{Luna2021}
{Luna}, M. \& {Moreno-Insertis}, F. 2021, \apj, 912, 75

\bibitem[{{Luna} {et~al.}(2017){Luna}, {Su}, {Schmieder}, {Chandra}, \&
  {Kucera}}]{Luna_Large-amplitude_2017}
{Luna}, M., {Su}, Y., {Schmieder}, B., {Chandra}, R., \& {Kucera}, T.~A. 2017,
  \apj, 850, 143

\bibitem[{{Luna} {et~al.}(2022){Luna}, {Terradas}, {Karpen}, \&
  {Ballester}}]{luna_extension_2022}
{Luna}, M., {Terradas}, J., {Karpen}, J., \& {Ballester}, J.~L. 2022, \aap,
  660, A54

\bibitem[{Luna {et~al.}(2016)Luna, Terradas, Khomenko, Collados, \&
  Vicente}]{luna_robustness_2016}
Luna, M., Terradas, J., Khomenko, E., Collados, M., \& Vicente, A.~d. 2016, The
  Astrophysical Journal, 817, 157, publisher: IOP Publishing

\bibitem[{Mazumder {et~al.}(2020)Mazumder, Pant, Luna, \&
  Banerjee}]{mazumder_simultaneous_2020}
Mazumder, R., Pant, V., Luna, M., \& Banerjee, D. 2020, Astronomy and
  Astrophysics, 633, A12

\bibitem[{{Moreton} \& {Ramsey}(1960)}]{Moreton1960b}
{Moreton}, G.~E. \& {Ramsey}, H.~E. 1960, \pasp, 72, 357

\bibitem[{{Morgan} \& {Druckm{\"u}ller}(2014)}]{Morgan2014}
{Morgan}, H. \& {Druckm{\"u}ller}, M. 2014, \solphys, 289, 2945

\bibitem[{{Okamoto} {et~al.}(2004){Okamoto}, {Nakai}, {Keiyama}, {Narukage},
  {UeNo}, {Kitai}, {Kurokawa}, \& {Shibata}}]{okamoto2004}
{Okamoto}, T.~J., {Nakai}, H., {Keiyama}, A., {et~al.} 2004, \apj, 608, 1124

\bibitem[{{Pant} {et~al.}(2016){Pant}, {Mazumder}, {Yuan}, {Banerjee},
  {Srivastava}, \& {Shen}}]{Pant2016}
{Pant}, V., {Mazumder}, R., {Yuan}, D., {et~al.} 2016, \solphys, 291, 3303

\bibitem[{Paraschiv {et~al.}(2022)Paraschiv, Donea, \&
  Judge}]{paraschiv_thermal_2022}
Paraschiv, A.~R., Donea, A.~C., \& Judge, P.~G. 2022, The Astrophysical
  Journal, 935, 172, publisher: The American Astronomical Society

\bibitem[{{Pesnell} {et~al.}(2012){Pesnell}, {Thompson}, \&
  {Chamberlin}}]{Pesnell2012}
{Pesnell}, W.~D., {Thompson}, B.~J., \& {Chamberlin}, P.~C. 2012, \solphys,
  275, 3

\bibitem[{Pouget(2007)}]{pouget07}
Pouget, G. 2007, {Analyse des protub{\'e}rances solaires observ{\'e}es {\`a}
  partir de la sonde solaire SOHO et du t{\'e}lescope Sacramento Peak:
  Oscillations, diagnostic, instabilit{\'e}s}

\bibitem[{Raes {et~al.}(2017)Raes, {Van Doorsselaere}, Baes, \&
  Wright}]{raes_observations_2017}
Raes, J.~O., {Van Doorsselaere}, T., Baes, M., \& Wright, A.~N. 2017, Astronomy
  \& Astrophysics, 602, A75

\bibitem[{Saqri {et~al.}(2020)Saqri, Veronig, Heinemann, Hofmeister, Temmer,
  Dissauer, \& Su}]{saqri_differential_2020}
Saqri, J., Veronig, A.~M., Heinemann, S.~G., {et~al.} 2020, Solar Physics, 295,
  6

\bibitem[{{Schmieder} {et~al.}(1985){Schmieder}, {Malherbe}, \&
  {Raadu}}]{Schmieder1985}
{Schmieder}, B., {Malherbe}, J.~M., \& {Raadu}, M.~A. 1985, \aap, 142, 249

\bibitem[{{Schou} {et~al.}(2012){Schou}, {Scherrer}, {Bush}, {Wachter},
  {Couvidat}, {Rabello-Soares}, {Bogart}, {Hoeksema}, {Liu}, {Duvall}, {Akin},
  {Allard}, {Miles}, {Rairden}, {Shine}, {Tarbell}, {Title}, {Wolfson},
  {Elmore}, {Norton}, \& {Tomczyk}}]{Schou2012}
{Schou}, J., {Scherrer}, P.~H., {Bush}, R.~I., {et~al.} 2012, \solphys, 275,
  229

\bibitem[{Schutgens \& Toth(1999)}]{schutgens_numerical_1999}
Schutgens, N. A.~J. \& Toth, G. 1999, Astronomy and Astrophysics, 345, 1038,
  publisher: EDP Sciences

\bibitem[{{Shen} {et~al.}(2014{\natexlab{a}}){Shen}, {Ichimoto}, {Ishii},
  {Tian}, {Zhao}, \& {Shibata}}]{shen2014a}
{Shen}, Y., {Ichimoto}, K., {Ishii}, T.~T., {et~al.} 2014{\natexlab{a}}, \apj,
  786, 151

\bibitem[{{Shen} {et~al.}(2014{\natexlab{b}}){Shen}, {Liu}, {Chen}, \&
  {Ichimoto}}]{shen2014b}
{Shen}, Y., {Liu}, Y.~D., {Chen}, P.~F., \& {Ichimoto}, K. 2014{\natexlab{b}},
  \apj, 795, 130

\bibitem[{Shibata \& Magara(2011)}]{shibata_solar_2011}
Shibata, K. \& Magara, T. 2011, Living Reviews in Solar Physics, 8, 6,
  publisher: Springer International Publishing

\bibitem[{{Sindhuja} {et~al.}(2022){Sindhuja}, {Singh}, {Asvestari}, \&
  {Raghavendra Prasad}}]{Sindhuja2022}
{Sindhuja}, G., {Singh}, J., {Asvestari}, E., \& {Raghavendra Prasad}, B. 2022,
  \apj, 925, 25

\bibitem[{{Sturrock}(1966)}]{Sturrock1966}
{Sturrock}, P.~A. 1966, \nat, 211, 695

\bibitem[{{Takahashi} {et~al.}(2015){Takahashi}, {Asai}, \&
  {Shibata}}]{takahashi2015}
{Takahashi}, T., {Asai}, A., \& {Shibata}, K. 2015, \apj, 801, 37

\bibitem[{{Vr{\v s}nak} {et~al.}(2007){Vr{\v s}nak}, {Veronig}, {Thalmann}, \&
  {{\v Z}ic}}]{vrsnak2007}
{Vr{\v s}nak}, B., {Veronig}, A.~M., {Thalmann}, J.~K., \& {{\v Z}ic}, T. 2007,
  \aap, 471, 295

\bibitem[{Wang {et~al.}(2016)Wang, Liu, Zimovets, Hu, Dai, \&
  Yang}]{wang_sympathetic_2016}
Wang, R., Liu, Y.~D., Zimovets, I., {et~al.} 2016, The Astrophysical Journal
  Letters, 827, L12, publisher: The American Astronomical Society

\bibitem[{{Wang} {et~al.}(2016){Wang}, {Zhang}, {Liu}, {Shen}, {Shen}, {Yang},
  {Zic}, {Vrsnak}, {Webb}, {Liu}, {Wang}, {Zhang}, {Hu}, \&
  {Zhuang}}]{Wang2016}
{Wang}, Y., {Zhang}, Q., {Liu}, J., {et~al.} 2016, Journal of Geophysical
  Research (Space Physics), 121, 7423

\bibitem[{{Xue} {et~al.}(2014){Xue}, {Yan}, {Qu}, \& {Zhao}}]{xue2014}
{Xue}, Z.~K., {Yan}, X.~L., {Qu}, Z.~Q., \& {Zhao}, L. 2014, in {Astronomical
  Society of the Pacific Conference Series}, Vol. 489, {Solar Polarization 7},
  ed. K.~N. {Nagendra}, J.~O. {Stenflo}, Q.~{Qu}, \& M.~{Samooprna}, 53

\bibitem[{Zhang {et~al.}(2019)Zhang, Fang, \& Chen}]{zhang_damping_2019}
Zhang, L.~Y., Fang, C., \& Chen, P.~F. 2019, The Astrophysical Journal, 884,
  74, publisher: American Astronomical Society

\bibitem[{{Zhang} {et~al.}(2012){Zhang}, {Chen}, {Guo}, {Fang}, \&
  {Ding}}]{Zhang2012}
{Zhang}, Q.~M., {Chen}, P.~F., {Guo}, Y., {Fang}, C., \& {Ding}, M.~D. 2012,
  \apj, 746, 19

\bibitem[{Zhang {et~al.}(2013)Zhang, Chen, Xia, Keppens, \&
  Ji}]{zhang_parametric_2013}
Zhang, Q.~M., Chen, P.~F., Xia, C., Keppens, R., \& Ji, H.~S. 2013, Astronomy
  and Astrophysics, 554, 124, publisher: EDP Sciences

\bibitem[{{Zhang} {et~al.}(2017){Zhang}, {Li}, \& {Ning}}]{zhang2017}
{Zhang}, Q.~M., {Li}, D., \& {Ning}, Z.~J. 2017, \apj, 851, 47

\bibitem[{Zhou {et~al.}(2018)Zhou, Xia, Keppens, Fang, \&
  Chen}]{zhou_three-dimensional_2018}
Zhou, Y.-H., Xia, C., Keppens, R., Fang, C., \& Chen, P.~F. 2018, The
  Astrophysical Journal, 856, 179, publisher: IOP Publishing

\bibitem[{{Zurbriggen} {et~al.}(2021){Zurbriggen}, {C{\'e}cere}, {Sieyra},
  {Krause}, {Costa}, \& {Gim{\'e}nez de Castro}}]{Zurbriggen2021}
{Zurbriggen}, E., {C{\'e}cere}, M., {Sieyra}, M.~V., {et~al.} 2021, arXiv
  e-prints, arXiv:2110.07687

\end{thebibliography}

\bibliographystyle{aa}

\end{document}